\documentclass[conference]{IEEEtran}
\IEEEoverridecommandlockouts
\usepackage{cite}
\usepackage{amsmath,amssymb,amsthm,amsfonts}
\usepackage{thmtools}
\usepackage{algorithmic}
\usepackage{graphicx}
\usepackage{textcomp}
\usepackage{xcolor}
\usepackage[colorlinks=true, citecolor=magenta, linkcolor=blue]{hyperref}
\usepackage{stfloats}
\usepackage{subcaption}
\usepackage[inline]{enumitem}
\usepackage{booktabs}
\usepackage{quoting}
\usepackage{caption}

\declaretheoremstyle[%
  headfont=\bfseries,%
  headpunct={:},%
  notefont=\normalfont\bfseries,%
  notebraces={--~}{},
    qed=$\blacksquare$,
]{definitionstyle}
\theoremstyle{definition}
\declaretheorem[style=definitionstyle,name=Definition]{defn}

\def\BibTeX{{\rm B\kern-.05em{\sc i\kern-.025em b}\kern-.08em
    T\kern-.1667em\lower.7ex\hbox{E}\kern-.125emX}}
\begin{document}

\title{A Hetero-functional Graph Structural Analysis of the American Multi-modal Energy System}

\author{\IEEEauthorblockN{Dakota J. Thompson}
\IEEEauthorblockA{\textit{Thayer School of Engineering at Dartmouth} \\
Hanover NH, United States \\
Dakota.J.Thompson.TH@Dartmouth.edu}
\and
\IEEEauthorblockN{Amro M. Farid}
\IEEEauthorblockA{\textit{Thayer School of Engineering at Dartmouth} \\
Hanover NH, United States \\
\textit{MIT Mechanical Engineering}\\
Cambridge MA, United States \\
amfarid@mit.edu}}

\maketitle

\begin{abstract}
As one of the most pressing challenges of the 21st century, global climate change demands a host of changes across four critical energy infrastructures: the electric grid, the natural gas system, the oil system, and the coal system.  Unfortunately, these four systems are often studied individually, and rarely together as integrated systems. Instead, holistic multi-energy system models can serve to improve the understanding of these interdependent systems and guide policies that shape the systems as they evolve into the future.  The NSF project entitled ``American Multi-Modal Energy System Synthetic \& Simulated Data (AMES-3D)" seeks to fill this void with an open-source, physically-informed,  structural and behavioral machine-learning model of the AMES.   To that end, this paper uses a GIS-data-driven, model-based system engineering approach to develop structural models of the American Multi-Modal Energy System (AMES).  This paper produces and reports the hetero-functional incidence tensor, hetero-functional adjacency matrix, and the formal graph adjacency matrix in terms of their statistics.  This work compares these four hetero-functional graph models across the states of New York (NY), California (CA), Texas (TX), and the United States of America (USA) as a whole.  From the reported statistics, the paper finds that the geography and the sustainable energy policies of these states are deeply reflected in the structure of their multi-energy infrastructure systems and impact the full USA's structure.
\end{abstract}


\section{Introduction} 
As one of the most pressing challenges of the 21st century, global climate change demands a host of changes across at least four critical energy infrastructures: the electric grid, the natural gas system, the oil system, and the coal system.  In the context of the United States,  this paper refers to this system-of-systems as ``the American Multi-Modal Energy System (AMES)".  The needs of climate change demand mitigation and adaptation strategies which are far more demanding than the needs of just mitigation alone.  Therefore, as policies are developed to drive the sustainable energy transition forward, they must not just aim to \emph{mitigate} climate change but must also \emph{adapt} to its effects with resilient architectures.  In effect, the need for decarbonization must be harmonized with the need for economic development, national energy security, and equitable energy access\cite{Moniz:2022:00,action:2010:00,wec:2016:00}.  These combined requirements to develop effective policies necessitate an understanding of the AMES interdependence's and how they vary geographically and temporally\cite{bridge:2013:00}.  Furthermore, this cross-sectoral interdependency can introduce architectural fragility\cite{Vespignani:2010:00} that must be managed as an integral part of the sustainable energy transition.  

Holistic multi-energy system models should serve to improve the understanding of these interdependent systems as they evolve into the future.  Unfortunately, while there have been many attempts at modeling multi-energy systems and large scale flows of energy, the field remains relatively nascent\cite{Albert:2004:00,Dong:2015:00,Jean-Baptiste:2003:00,Kriegler:2018:00,Robert-Lempert:2019:00,Rogers:2013:00,Lara:2020:00,Panteli:2019:00,Mejia-Giraldo:2012:00}.  While initial models were developed in response to the oil crisis and then to mitigate climate change with decarbonization they have also been developed to introduce new energy streams such as hydrogen, synthetic fuels, bio-fuels, and other renewable energy sources.  These works introduce their own set of weaknesses including the lack of asset level granularity, difficulty of use, and specific one-off geographically-specific use case models.  Additionally, a majority of the works investigating these energy systems in the past have been performed on individual energy networks \cite{masters:2013:00,mokhatab:2012:00,Lurie:2009:00,EIA:2014:11,Priyanka:2021:00}. More recently, work has been published analyzing only a couple systems together such as pairing the electric grid with one of the others fossil fuel systems that compose the AMES\cite{Albert:2004:00,Dong:2015:00,Jean-Baptiste:2003:00,Kriegler:2018:00,Robert-Lempert:2019:00,Rogers:2013:00,Lara:2020:00,Panteli:2019:00}.  These works however, do not include all four critical energy infrastructures and do not extend to the entire American geography.  As an exception, the EIA developed a comprehensive model called the National Energy Modeling System (NEMS) which it uses to produce the (American) Annual Energy Outlook\cite{EIA:2020:00}.  Despite serving this important function and being publicly available, this software tool remains opaque and difficult to use.  The EIA website itself recognizes: "[The] NEMS is only used by a few organizations outside of the EIA. Most people who have requested NEMS in the past have found out that it was too difficult or rigid to use \cite{EIA:2017:01}".  Consequently, holistic multi-energy system models of the AMES remain a present need for open-source, citizen-science to inform policies.

With a deficit of spatially and functionally resolved data, and with the current methods for modeling multi-energy systems having their limitations, the (American) National Science Foundation (NSF) put forth a call for ``research to develop and make available simulated and synthetic data on interdependent critical infrastructures (ICIs), and thus to improve understanding and performance of these systems"\cite{Johnson:2017:00}.  The NSF project entitled "American Multi-Modal Energy System Synthetic \& Simulated Data (AMES-3D)" seeks to fill this void with an open-source structural and behavioral model of the AMES.  Adhering to a Model Based Systems Engineering (MBSE) approach \cite{Dori:2015:00,Friedenthal:2011:00}, this project develops an interdependent system data set and its associated models on top of a strong theoretical foundation in systems engineering.  As a result, it can be used for practical applications in the energy systems field to address not just mitigation of climate change but adaptation and resilience as well.  This is made possible by using asset-level, openly-available datasets to infer the AMES' reference architecture\cite{Thompson:2020:00} to organize and define the interconnections between the four subsystems.  The reference architecture uses SysML\cite{Friedenthal:2011:00} to model the four interdependent energy systems, and the flows of mass and energy within and between them.  This reference architecture provides a more detailed and self-consistent MBSE foundation for energy models moving forward relatively to the ``reference energy systems" that have been used in some national energy system optimization models\cite{PlazasNino:2022:00, Millot:2020:00, Dedinec:2016:00,Yangka:2016:00}. The datasets used to infer reference architecture are also used to instantiate the AMES into an instantiated architecture\cite{Thompson:2020:00}.  While the NSF project seeks to develop both a structural and behavioral model, this paper focuses its scope on the former.  

The development of the AMES reference architecture provides several immediate benefits.  The first is that a SysML based reference architecture describes the system's form, function and the allocation of the latter onto the former\cite{Dori:2015:00,Friedenthal:2011:00}.  Therefore, the reference architecture describes not just what the system is made of, but also what it does.  Second, a SysML based reference architecture can be readily translated into mathematical models including both the form and function. Standard structural models include formal graphs that describe energy facilities and how they are connected.  In the meantime, hetero-functional graphs (HFG) describe how the wide variety of capabilities in a system are interconnected and the flows of operands between them.  HFGs have been shown to provide more information than formal graphs when analyzing an evolving instantiated architecture\cite{Thompson:2020:01, Thompson:2021:00}.  In effect, HFGT provides a means to quantitatively interpret the graphical SysML-based models from both a formal and functional lens.  Such an analysis has already been conducted on a small scale electric power distribution systems \cite{Thompson:2020:01} as well as on a large scale for the entirety of the American electric power system\cite{Thompson:2021:00}.  This paper now builds on these electricity-only analyses to study the structure of the whole AMES.

\subsection{Original Contributions}
This paper uses a data-driven, MBSE-guided approach to develop open-source structural models of the American Multi-Modal Energy System.  More specifically, the AMES reference architecture\cite{Thompson:2020:00} is applied to an asset-level GIS data called Platts Map Data Pro \cite{Platts:2017:00} to create models of several regions.  The instantiated structural models include for the first time the electric grid, the natural gas system, the oil system, the coal system and the interconnections between them as defined by the AMES reference architecture, for the full contiguous United States of America (USA).  Initial results are organized into two categories; a formal and hetero-functional graph for each of the regions being studied, New York (NY), California (CA), Texas (TX), and the full contiguous (USA).  The states are chosen for their large size and the diversity of their energy policies.  Consequently, the chosen regions have also taken distinct directions to advance the sustainable energy transition.  In 2019, CA had the most renewable energy generation out of all the states \cite{EIA:2021:01}.  In the meantime, NY's efforts to expand renewable energy capacity are balanced by its reliance on natural gas and oil to meet space heating energy demands\cite{EIA:2021:00}.  Alternatively, Texas, while being the nations leading crude oil and natural gas producing state, is also the nations leading producer of wind powered electric generation\cite{EIA:2022:00}. By using MBSE and HFGT, new open source data models are presented for these three states and the full USA to aid in advancing and guiding the sustainable energy transition and energy policies.  

\subsection{Paper Outline}
The remainder of the paper proceeds as follows.  Section \ref{Background} is a description of the background literature and the lack of open data models used to develop the instantiated architecture models and guide policy.  Section \ref{methodology} then presents the data source that drove the instantiated models in subsection \ref{InputData}, followed by defining the AMES Reference Architecture in subsection \ref{AMESRef}.  The data processing in subsection \ref{AMESRef} in then presented followed by a subsection on hetero-functional graph theory\ref{HFGTToolbox}. The paper then presents a comparison of the formal graphs and hetero-functional graphs network statistics for each state in the Results section \ref{Results}.  The Results first starts with an analysis of the computational intensity in section \ref{CompInt}, it then presents the formal and hetero-functional graph statistics in sections \ref{FGStats} and \ref{HFGStats} respectfully.  Section \ref{Results} continues to then present the formal graph resource distributions in section \ref{FGResDist} and the HFG capability distributions in section \ref{HFGCapDist}.  A discussion of the HFG process degree distributions in section \ref{HFGProcDist} brings the methodology section to close. Finally the paper is brought to a conclusion in section \ref{Conclusions}. 

\section{Background}\label{Background}
This section serves to situate the paper with respect to existing literature and the ongoing trends in the field.  Section \ref{Sec:RefArch} describes existing multi-energy system models.  Section \ref{Sec:IIS} discusses the general lack of interdependent infrastructure system data and models.  Section \ref{Sec:Transparency} discusses the emergy trend towards open data and source code in the multi-energy system field.  Finally, Section \ref{Sec:HFGT} introduces some essential concepts in HFGT that are used in the remainder of the paper.    
\subsection{Existing Multi-Energy System Models}\label{Sec:RefArch}
Many existing multi-energy system models are in effect national models used to inform national energy policy.   Many such national models have an often implicitly stated ``reference energy system" which serves as the first step for defining energy flow relationships\cite{PlazasNino:2022:00, Millot:2020:00, Dedinec:2016:00,Yangka:2016:00}.  

\begin{defn}\textbf{- Reference Energy System}\cite{PlazasNino:2022:00} As a tool to begin modeling energy systems, \textit{``it represents a simplified and aggregated graphical representation of the entire energy system under analysis which shows all existing and potential new energy supply chains from primary energy to final demand."}
\end{defn}

A reference energy system defines how the primary energy flows are processed and converted into different energy carriers and where the energy is ultimately utilized in end-use sectors.  While the RES does identify all the modes of energy being tracked through the model, it is driven purely by system behavior.  This means it only incorporates the system function but does not explicitly describe system form.  In contrast, the practice of MBSE requires the definition of system function and system form\cite{Dori:2015:00,Friedenthal:2011:00} in a graphical modeling language like SySML.  When a ``reference energy system" omits the system form, asset-level, and technology-specific interactions are ultimately neglected.  Additionally, by ignoring the form at the starting point for these energy system models, simulation model outputs and results become less transparent.  Consequently, the absence of a description of system form further impedes the comparability of different models because they do not explicitly state which technologies exist and how they interact.  This lack of model comparability further impedes the comparability of differing strategies to global climate change mitigation and adaptation.  As an alternative to RES models,  MBSE and SysML utilize reference architectures as the basis of all downstream modeling activities.  

\begin{defn}\textbf{- Reference Architecture\cite{Cloutier:2010:00}} \textit{``The reference architecture captures the essence of existing architectures, and the vision of future needs and evolution to provide guidance to assist in developing new instantiated system architectures. ...Such reference architecture facilitates a shared understanding across multiple products, organizations, or disciplines about the current architecture and the vision on the future direction. A reference architecture is based on concepts proven in practice. Most often preceding architectures are mined for these proven concepts. For architecture renovation and innovation validation and proof can be based on reference implementations and prototyping. In conclusion, the reference architecture generalizes instantiated system architectures to define an architecture that is generally applicable in a discipline. The reference architecture does however not generalize beyond its discipline."}
\end{defn}

A reference architecture specifically includes a description of system form and function.   Consequently, it makes transparent all flows of matter and energy and which energy assets are used for transforming and transporting these flows. Furthermore, It becomes much easier to compare energy system models and their underlying assumptions.  Finally, it is straightforward to determine how much of a reference architecture appears in the instantiated architecture that pertains to the geography of a specific region or case study.  

\begin{defn}\textbf{- Instantiated Architecture}\cite{Cloutier:2010:00,Friedenthal:2011:00} A case specific architecture, which represents a real-world scenario, or an example test case. At this level, the physical architecture consists of a set of instantiated resources, and the functional architecture consists of a set of instantiated system processes. The mapping defines which resources perform what processes.
\end{defn}

The transparency and comparability of a reference architecture becomes even more valuable in light of the open energy modeling initiative.  It began in 2014 as a result of the open access movement which had begun in 2010\cite{Oberle:2019:00,PlazasNino:2022:00}.  These movements originated from a recognition of the limited transparency and reproducibility of energy systems  models\cite{Lopion:2018:00}.  The open energy modeling initiative seeks to encourage the use of open access models in research to guide energy policies and the sustainable energy transition.  While this initiative has resulted in the production of many open access models, much work remains to compare and converge these models towards the ultimate goals of sustainable energy transition. 


Several powerful energy system models have come out of the energy modeling initiative including the OSeMOSYS, NEMS, and PRIMES\cite{PlazasNino:2022:00, HOWELLS:2011:00, EIA:2019:00,E3MLAB:2018:00}.  However, these models each had their own weaknesses.  For example, as previously mentioned, the NEMS is extremely difficult to use making it unusable for many open access projects.  While the NEMS was a large model incorporating many different modes of energy to inform the annual energy outlook for policy planning,  there have also been a plethora of models developed for specific systems rather than focusing on the interconnection of multiple critical infrastructures\cite{masters:2013:00,mokhatab:2012:00,Lurie:2009:00,EIA:2014:11,Priyanka:2021:00}.  Alternatively, other energy system optimization models (ESOMs) such as the EnergyScope TD have been designed for the analysis of intermittent integration of renewable on an hourly scale.  Similarly, the Electric Power Enterprise Control Simulator addresses the integration of variable renewable energy on multiple time-scales\cite{Muzhikyan:2018:SPG-W07,Muhanji:2020:EWN-J44}.  These operational time-scale models, however, are not designed to explore the annual transformation pathways of the sustainable energy transition\cite{PlazasNino:2022:00, Stralen:2021:00, Limpens:2019:00}.  Finally, the development of multi-energy system models did not truly gain concerted attention until 2016\cite{Heendeniya:2020:00}.  

\subsection{Interdependent Infrastructure System Data \& Models}\label{Sec:IIS}

As multi-energy system model developed, it created a greater need for interdependent infrastructure system data and models.  In 2016, the NSF released a call for open interdependent critical infrastructure system data\cite{Johnson:2017:00} that could be specifically used in system resilience and climate change adaptation research.  This call directly addresses the lack of existing open data sources on energy systems.  

In the meantime, the existing methods for organizing such data into multi-modal energy system models had its own limitations.  Multi-layer networks, for example, was often looked at as the leading candidate to structurally model these interdependent infrastructures.  Unfortunately, multi-layer networks have often been unable to address the explicit heterogeneity often encountered in engineering systems\cite{Schoonenberg:2019:ISC-BK04,Kivela:2014:00}.  In a recent comprehensive review, Kivela et.al \cite{Kivela:2014:00} wrote: 
\begin{quoting}
``The study of multi-layer networks $\ldots$ has become extremely popular.  Most real and engineered systems include multiple subsystems and layers of connectivity and developing a deep understanding of multi-layer systems necessitates generalizing `traditional' graph theory.  Ignoring such information can yield misleading results, so new tools need to be developed.  One can have a lot of fun studying `bigger and better' versions of the diagnostics, models and dynamical processes that we know and presumably love -- and it is very important to do so but the new `degrees of freedom' in multi-layer systems also yield new phenomena that cannot occur in single-layer systems.  Moreover, the increasing availability of empirical data for fundamentally multi-layer systems amidst the current data deluge also makes it possible to develop and validate increasingly general frameworks for the study of networks.  

$\ldots$ Numerous similar ideas have been developed in parallel, and the literature on multi-layer networks has rapidly become extremely messy.  Despite a wealth of antecedent ideas in subjects like sociology and engineering, many aspects of the theory of multi-layer networks remain immature, and the rapid onslaught of papers on various types of multilayer networks necessitates an attempt to unify the various disparate threads and to discern their similarities and differences in as precise a manner as possible.

$\ldots$ [The multi-layer network community] has produced an equally immense explosion of disparate terminology, and the lack of consensus (or even generally accepted) set of terminology and mathematical framework for studying is extremely problematic."
\end{quoting}

In this context, the NSF funded the "American Multi-Modal Energy System Synthetic \& Simulated Data (AMES-3D)" to develop an open-source energy system data set using modeling methods, and more specifically hetero-functional graph theory (HFGT) that do not exhibit the limitations of multi-layer networks\cite{Schoonenberg:2019:ISC-BK04}.  Consequently, this paper uses HFGT to present a hetero-functional graph structural model (i.e. instantiated architecture) and analysis of the American multi-modal energy system as part of the NSF-funded AMES-3D project.

\subsection{Transparency through Open Data and Source Code}\label{Sec:Transparency}

In addition for the need for reference architectures (Sec. \ref{Sec:RefArch}), and the need for interdependent infrastructure system data and models (Sec. \ref{Sec:IIS}), the multi-energy systems literature also recognizes a pressing need for open data and source code.  To that effect,  the open modeling initiative (OMI) was founded in 2014\cite{Oberle:2019:00}.  It was founded with the intent to promote open access models research to guide energy policy\cite{PlazasNino:2022:00}.  Similarly, the  open modeling foundation (OMF) was founded in 2021\cite{PlazasNino:2022:00,OMF:2022:00} as an ``international open science community that works to enable the next generation modeling of human and natural systems" by making models more easily discoverable and globally accessible\cite{OMF:2022:00}.  The OMF seeks to create a collection of reusable, interoperable models to study complex interactions between people and the environment at multiple scales.  

In direct alignment with the missions of the OMI and OMF is the NSF's call for open interdependent critical infrastructure system data as previously mentioned.  The work presented in this paper addresses these needs with transparent open-source code and reproducible mathematics.  First, the work relies on the previously developed AMES reference architecture\cite{Thompson:2020:00} that serves to visualize all of the AMES components, functions, and interactions.  In many fields,  ``mature" reference architectures often develop into international standards that converge and reconcile the otherwise inevitable proliferation of models.  Second, the mathematics of hetero-functional graph theory is used to instantiate the AMES reference architecture for NY, CA, TX, and the contiguous United States.  The explicit statement of the HFGT mathematics not only validates the work and makes it entirely reproducible, but it also makes transparent the open-source HFGT toolbox used to produce the computational results of this work.  In all, the open data and open source commitments made in the AMES-3D project has necessitated a research methodology that is enhances usability, re-usability, transparency, and comparability of the models, tools, and results.

\subsection{Hetero-functional Graph Theory}\label{Sec:HFGT}
In the context of this work, hetero-functional graph theory serves to translate and instantiate a graphical, SysML-based reference architecture into its mathematically equivalent hetero-functional graph.  The reliance on a reference architecture grounds the work in a strong MBSE foundation.  Additional, the AMES reference architecture itself provides a clear definition of the assets, functionality, and modes of energy that are included in the AMES.  This lossless translation from an MBSE-SysML model to a HFG maintains the allocation of function onto form as capabilities composed of subject + verb + object sentences where the subjects are resources, the predicates are the processes, and the operands are the objects of the verbs.  This translation form the AMES reference architecture to a HFG (using the HFGT toolbxo) has been well demonstrated for the single-commodity American electric power system\cite{Thompson:2021:00}.  Now this demonstrates the same process to create an multi-energy instantiated model of the full continguous USA for a structural (network) analysis.  

The following HFGT definitions are presented to facilitate the remainder of the paper.

%
\begin{defn}[System Operand]\cite{SE-Handbook-Working-Group:2015:00}
An asset or object $l_i \in L$ that is operated on or consumed during the execution of a process.  
\end{defn}
\begin{defn}[System Process\cite{Hoyle:1998:00,SE-Handbook-Working-Group:2015:00}]\label{def:CH4:process}
An activity $p \in P$ that transforms a predefined set of input operands into a predefined set of outputs. 
\end{defn}
\begin{defn}[System Resource]\cite{SE-Handbook-Working-Group:2015:00}
An asset or object $r_v \in R$ that is utilized during the execution of a process.  
\end{defn}
\begin{defn}[Buffer\cite{Schoonenberg:2019:ISC-BK04}]\label{defn:BS}
    A resource $r \in R$ is a buffer $b_s \in B_S$ iff it is capable of storing one or more operands at a unique location in space.  $B_S=M \cup B$. 
\end{defn}
\begin{defn}[Capability\cite{Farid:2006:IEM-C02,Farid:2007:IEM-TP00,Farid:2008:IEM-J05,Farid:2008:IEM-J04,Farid:2015:ISC-J19,Farid:2016:ISC-BC06}]\label{defn:capability}
An action $e_{wv} \in {\cal E}_S$ (in the SysML sense) defined by a system process $p_w \in P$ being executed by a resource $r_v \in R$.  It constitutes a subject + verb + operand sentence of the form: ``Resource $r_v$ does process $p_w$".  
\end{defn}

\section{Methodology}\label{methodology}

As mentioned in the introduction, this paper utilizes a data driven, MBSE-guided methodology to infer a hetero-functional graph structural model of the American Multi-modal Energy System.  This section succinctly relays this method in four subsections:
\begin{enumerate}
\item Input Data:  Platts Map Data Pro
\item Infer AMES Reference Architecture
\item Convert GIS Shape Files to XML Data
\item Calculate Hetero-functional Graph Structural Model
\end{enumerate}

\begin{figure}[h!]
\vspace{-0.1in}
\includegraphics[width=3.25in]{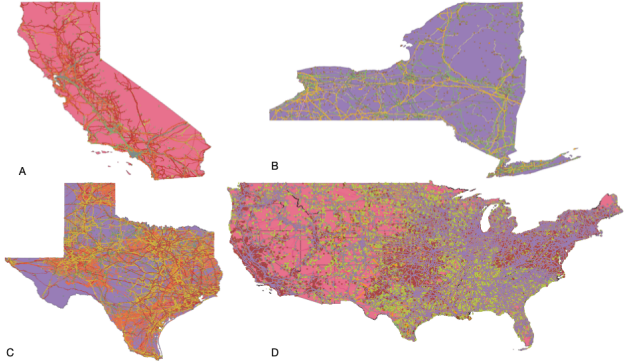} 
\caption{GIS Layers from the Platts Map Data Pro dataset for the electric grid, natural gas system, oil system, and coal system for California (A), New York (B), Texas (C), and the USA (D).}
\vspace{-0.1in}
\label{Fig:GIS}
\end{figure}

\subsection{Input Data: Platts Map Data Pro}\label{InputData}
Following a data driven approach, the Platts Map Data Pro \cite{Platts:2017:00} data set is used to infer and instantiate the AMES Structural models.  This input dataset consists of Graphic Information System (GIS) layers for each of the four subsystems in the AMES\cite{Platts:2017:00}.  These geo-spacial layers include meta-data attributes of the physical resources/facilities that compose the AMES infrastructure in addition to their GPS coordinates.  As the Platts Map Data Pro is directed towards wholesale energy decisions, the data is limited to transmission system resources and neglects distribution level assets.  This limitation in the dataset notably excludes retail distribution of oil and gas (by truck).   It also cuts the electric grid off at the substations excluding distributed electric generation assets such as roof-top solar that are an integral part of the sustainable energy transition. Nevertheless, the Platts Map Data Pro is likely the best available dataset because it allows inferences of not just the AMES's form but its function too.  Visualizations of the Platts map data pro GIS layers for each state addressed in this paper and the full USA are presented in Figure \ref{Fig:GIS}.

\begin{figure*}[!hb]
\centering
\includegraphics[width=6.5in]{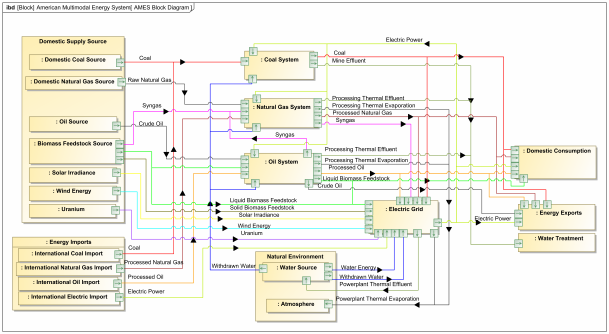}
\caption{The top level internal block diagram of the AMES.  The domestic supply sources, the energy imports, natural environment, domestic consumption, energy exports, and water treatment are external to the AMES four subsystems of coal, natural gas, oil, and electric grid.  }
\label{Fig:AMESRef}
\vspace{-0.2in}
\end{figure*}

\subsection{Infer AMES Reference Architecture}\label{AMESRef}
The inference of a hetero-functional graph structural model of the American Multi-modal Energy System first requires the inference of the underlying reference architecture.  In the data-driven steps that follow, the reference architecture plays a critical role in organizing the cleaned Platts GIS data into defined resources with proper allocated functionality.  Figure \ref{Fig:AMESRef} shows the top-level context diagram of the AMES reference architecture and it is further elaborated in prior work\cite{Thompson:2020:00}.  The AMES reference architecture provides a consistent blue print from which to develop AMES models irrespective of the choice of region or scale.  It also defines all energy resources/facilities, and the functions that they can perform.  The set of operands used to track the flows of mass and energy between its many functions are also defined.  Consequently, the AMES reference architecture supports downsteam analysis that address both climate change mitigation and adaptation.  

\subsection{Convert GIS Shape Files to XML Data}\label{GIS2XML}
In this step, the Platts Map Data Pro GIS shape files are converted into an associated XML file for each region (i.e. NY, CA, TX, USA) that serves as the input for the openly available HFGT toolbox \cite{Thompson:2020:02}.  Although the Platts Map Data Pro GIS shape files are a commercially-curated dataset, they still need substantial cleaning and processing before being organized into an XML file.  As an immediate first step, all resources marked with a canceled status, closed status, or illegible identifying attributes are removed.  From there, the primary obstacle is that the Platts GIS data does not guarantee physical continuity between all of the point-type (e.g. refineries and electric power generation facilities) and line-type (e.g. oil pipelines and electric power lines) resources in the dataset.  

A novel operand-guided geographical clustering algorithm is developed and then applied to all four energy subsystems simultaneously.  It ensures that point-type resources with overlapping GIS coordinates are aggregated into a single resource cluster.  It also ensures that line-type resources whose endpoints are not connected to any resource cluster (e.g. ``power lines to nowhere") are connected to a nearby resource cluster making sure to maintain the operand compatibility of the line-type resource to the point-type resource cluster.  For example, an oil pipeline must connect to a refinery rather than an electric power substation. 
 The clustering algorithm also removes any isolated point-resources from the dataset.  The geographical clustering algorithm provides an automated means for cleaning the Platts GIS data at the formidable scale presented by the AMES.     

The clustering algorithm proceeds in several steps.  
\begin{enumerate*}
\item Using a primary clustering distance of 0.1 miles, point-type resources and the endpoints of line-type resources  endpoints with like operand associations are sorted into clusters and connected in a manner that respects the physical continuity of operands.
\item Then with a secondary clustering distance of 1 mile, the remaining isolated nodes are added to existing clusters with like operands.
\item Finally, with a tertiary clustering distance of 35 miles,  the remaining isolated nodes are connected to existing clusters with like operands via the creation of new transportation resources.
\end{enumerate*}
After the execution of this clustering algorithm, the data attributes of these resources are converted to strings in a manner that adheres to the AMES reference architecture and the XML file format required by the hetero-functional graph theory toolbox\cite{Thompson:2020:02}.  

\subsection{Calculation of Hetero-functional Graph Structural Model}\label{HFGTToolbox}
The next step is to run the HFGT toolbox\cite{Thompson:2020:02} on this newly produced XML file so as to produce the positive and negative hetero-functional incidence tensors (HFITs).  
\begin{defn}[The Negative 3$^{rd}$ Order Hetero-functional Incidence Tensor (HFIT) $\widetilde{\cal M}_\rho^-$]\cite{farid:2021:00,Thompson:2022:ISC-C80}
The negative hetero-functional incidence tensor $\widetilde{\cal M_\rho}^- \in \{0,1\}^{|L|\times |B_S| \times |{\cal E}_S|}$  is a third-order tensor whose element $\widetilde{\cal M}_\rho^{-}(i,y,\psi)=1$ when the system capability ${\epsilon}_\psi \in {\cal E}_S$ pulls operand $l_i \in L$ from buffer $b_{s_y} \in B_S$.
\end{defn} 

\begin{defn}[The Positive  3$^{rd}$ Order Hetero-functional Incidence Tensor (HFIT)$\widetilde{\cal M}_\rho^+$]\cite{farid:2021:00,Thompson:2022:ISC-C80}
The positive hetero-functional incidence tensor $\widetilde{\cal M}_\rho^+ \in \{0,1\}^{|L|\times |B_S| \times |{\cal E}_S|}$  is a third-order tensor whose element $\widetilde{\cal M}_\rho^{+}(i,y,\psi)=1$ when the system capability ${\epsilon}_\psi \in {\cal E}_S$ injects operand $l_i \in L$ into buffer $b_{s_y} \in B_S$.
\end{defn} 

\noindent In the context of the AMES, the operands are flows of matter and energy like coal, oil, natural gas, and electricity.  The buffers are the point-facilities like electric power plants and refineries.  The capabilities are ``subject + verb + object" sentences such as ``NG refinery refines raw natural gas".  The HFITs are important because they include all of the information necessary to produce:
\begin{enumerate*}
\item a Formal Graph (FG) adjacency matrix $(A_{Bs})$ where point-facilities are connected via edge-facilities, and
\item a Hetero-Functional Graph (HFG) adjacency matrix $A_\rho$ where the system's capabilities follow one another.  
\end{enumerate*}

\begin{defn}[The Formal Graph Adjacency Matrix $A_{Bs}$] The formal graph adjacency matrix  ${A_{Bs} \in \{0,1\}^{|B_s|\times |B_S|}}$  is a matrix whose element ${A_{Bs}(y_1,y_2)=1}$ when there is a physical connection between buffer $y_1$ and buffer $y_2$.
\end{defn}
\begin{defn}[Hetero-functional Adjacency Matrix\cite{Farid:2015:ISC-J19}]\label{def:HFAM}
A square binary matrix $A_\rho$ of size $|{\cal E}_S|\times |{\cal E}_S|$ whose element $A_\rho(\psi_1,\psi_2)\in \{0,1\}$ is equal to one when string $z_{\psi_1,\psi_2} \in {\cal Z}$ is available and exists.
\end{defn}

\begin{table*}[b!]
\vspace{-0.1in}
\caption{The computational complexity of processing GIS data and the HFGT Toolbox for the Multi-Modal Energy Systems in NY, CA, TX, and the USA.}\label{Ta:HFGC}
\vspace{-0.1in}
\begin{center}
\begin{tabular}{l c c c c}
& New York & California & Texas & United States\\\hline
Time to Assign Clusters (sec)            & 2.44  & 9.9   & 166.49 & 3,963.55\\\hline  
Time to Process Data (sec)            & 23.60 & 105.31 & 1,445.81 & 49,373.63\\\hline
Time to Process Data \& Write XML (sec)   & 28.29 & 121.43 & 1,499.93 & 51,470.68\\\hline
Time to Calculate $A_\rho$ (sec)          & 14.49 & 20.25 & 1,252.30 & 6,171.14\\\hline
Time to run HFGT Toolbox (sec)            & 19.13 & 29.45 & 1,382.97 & 6,439.74\\\hline
XML size (MB)           & 5.6 &  13 &  192 & 420.7\\\hline
HFG resulting Pickle size (MB) & 8.8 & 20.6 & 314.9 & 682.6\\
\bottomrule
\end{tabular}
\end{center}
\vspace{-0.1in}
\end{table*}

Each of these adjacency matrices are then calculated from the incidence tensors.  The formal graph adjacency matrix $A_{Bs}$ requires two steps.  First, the two HFITs are summed along the operand dimension to produce two incidence matrices\cite{farid:2021:00,Thompson:2022:ISC-C80}:

\vspace{-0.2in}
\begin{align}\label{Eq:FG1}
M_B^{+}(y,\psi) =& \sum_i^{|L|} {\cal M}^{+}_\rho(i,y,\psi)\\\label{Eq:FG2}
M_B^{-}(y,\psi) =& \sum_i^{|L|} {\cal M}^{-}_\rho(i,y,\psi)
\end{align}
\noindent It is important to recognize that the operand heterogeneity information is lost in this process. Then, these incidence matrices are multiplied.  This multiplication also results in a loss of information where the allocation of function onto form in the form of capabilities is dropped to describe the physical connection of buffers\cite{farid:2021:00,Thompson:2022:ISC-C80}.
\begin{align}\label{Eqn:IM2FGAM}
A_{Bs} = M_B^+M_B^{-T}
\end{align}
\noindent In the meantime, the hetero-functional graph adjacency matrix $A_\rho$ is calculated without loss of information after the HFITs have been matricized (or flattened) into hetero-functional incidence matrices $M_\rho^{+}$ and $M_\rho^-$ with dimension $|L||B_S|\times |{\cal E}_S|$\cite{farid:2021:00,Thompson:2022:ISC-C80}.  

\vspace{-0.2in}
\begin{align}\label{Eqn:IM2HFAM}
 A_\rho = M_\rho^{+T}M_\rho^-
\end{align}

While the FG adjacency matrix shows the physical connections from one point-resource (i.e. buffer) to another, the HFG adjacency matrix shows the logical sequence of capabilities one after the other.  This results in the HFG adjacency matrix not just describing the physical connections but also the flow of functionality on matter and energy.  As previous works have shown, the HFG allows for more comprehensive resilience analyses; be it for small electric power distributions systems \cite{Thompson:2020:01} or for full scale analysis of the American electric power system \cite{Thompson:2021:00}.  The open-source HFGT toolbox \cite{Thompson:2020:02} provides an automated means for processing the input XML files at the formidable scale presented by the AMES.     

\section{Results:  Structural Model Statistics}\label{Results}
Once created, the hetero-functional incidence tensor, the formal graph, and the hetero-functional graph for the four regions of NY, CA, TX, and the full USA can be compared.  The following section presents these comparisons in the following subsections:

\begin{enumerate}
\item Computational Performance of the HFGT Toolbox
\item Formal Graph Statistics
\item Hetero-Functional Graph Statistics
\item Formal Graph Resource Distribution
\item Hetero-Functional Capability Distribution 
\item Hetero-Functional Graph Process and Operand Degree Distribution
\end{enumerate}

\subsection{Computational Performance of HFGT Toolbox}\label{CompInt}

The hetero-functional graph theory toolbox\cite{Thompson:2020:01} is used as a computational tool that supports hetero-functional graph theory computations\cite{Schoonenberg:2019:ISC-BK04,farid:2021:00,Farid:2006:IEM-C02}.  For this work, the exceptional scale of the AMES required extensive computational improvements to the HFGT toolbox so as to reduce computation times and memory used to produce the XML files and HFGs.  For this work, all the models and data processing to produce the XML files was done on an iMac desktop with a 4 GHz Quad-Core Intel Core i7 processor and 32GBs of RAM.  The computation times of major code milestones are presented in Table \ref{Ta:HFGC}.   The resulting file sizes are also presented in Table \ref{Ta:HFGC}.  Note that the overwhelming majority of the computation is devoted to converting the GIS shape files into XML files. More specifically, automatic data cleaning and processing activities dominated the computation. In comparison, 

Interestingly, the majority of the computation time came from processing and cleaning the GIS data.  Specifically, the processing and cleaning post assigning clusters to points takes the largest amount of time when producing the XML.  The actual assignment of clusters takes about $1/10^{th}$ of the total time to produce the XMLs for all regions. When looking at the computation time of the HFGT Toolbox nearly all of the required time is spent calculating $A_{\rho}$.  This calculation is ported over to Julia from Python through the use of CSV files, $A_{\rho}$ is calculated, then ported back to Python via another CSV.  The hetero-functional adjacency matrix is at the heart of the HFG as it does represent all flows of capabilities with the system.  In all, the HFGT toolbox computations themselves are highly optimized and are produced in under two hours on a moderately sized computer even for the full AMES dataset.  

In line with the scale of each region, the smaller the region is, the faster both the XML file and HFG is produced; with NY being the fastest followed by CA, TX, and then the full USA.  While CA is about twice the size of NY it took 4 times as long to produce the XML than NY.  When comparing TX to NY it can be seen that TX is 26 times as large and takes 60 times as long to produce an XML.  This suggests the computation efficiency of creating an XML is that of $2N$.  When comparing the computation times for the HFGT Toolbox, the same trend emerges suggesting the toolbox also has a computational efficiency of $2N$. Additionally, seeing that TX takes about a quarter of the time to complete the HFGT toolbox as the USA, it suggests that TX composes about half of the American energy infrastructure.  The significant portion of energy resources that TX commands is confirmed by the later sections.




\subsection{Formal Graph Statistics}\label{FGStats}

From the assessment of the computational intensity, the basic statistics of the formal graph models for the four regions can be compared relative to the size of their populations and land areas (Table \ref{Ta:FGS}).

\begin{table*}[t!]
\vspace{-0.1in}
\caption{Formal Graph Statistics for the Multi-Modal Energy Systems in NY, CA, TX, and the USA.}\label{Ta:FGS}
\vspace{-0.1in}
\begin{center}
\begin{tabular}{l c c c c}\toprule
& New York & California & Texas & United States\\\hline
\# of Operands                      & 13 & 13 & 14 & 14\\\hline  
\# of Buffers                       & 7,686 & 16,754 & 197,108 & 473,321\\\hline
\# of Edges                         & 9,115 & 20,674 & 179,895 & 511,802\\\hline
Formal Graph Sparsity               & 1.5438e-4 & 7.3661e-5 & 4.6306e-6 & 2.2848e-6\\\hline
Population (millions)               & 19.8 & 39.2 & 29.5 & 306.7\\\hline
Land Area (sqr miles)               & 54,556 & 163,696 & 268,597 & 3,119,884\\\hline
Population Density(ppl/sqr mile)    & 362.9 & 239.5 & 109.8 & 98.3\\\hline
Buffers/Land Density (buffers/sqr mile) & 0.1409 & 0.1023 & 0.7338 & 0.1518\\\hline
Buffers/Population Density (buffers/ppl) & 3.8818e-4 & 4.2740e-4 & 6.6816e-3 & 1.5432e-3\\\hline 
Edges/Land Density (edges/sqr mile) & 0.1671 & 0.1263 & 0.6698 & 0.1641\\\hline
Edges/Population Density (edges/ppl) & 4.6035e-4 & 5.2740e-4 & 6.0981e-3 & 1.6687e-3\\\bottomrule
\end{tabular}
\end{center}
\vspace{-0.2in}
\end{table*}

\begin{table*}[b!]
\vspace{-0.1in}
\caption{Hetero-functional Incidence Tensor Statistics for the Multi-Modal Energy Systems in NY, CA, TX, and the USA.}\label{Ta:HFGS}
\vspace{-0.1in}
\begin{center}
\begin{tabular}{l c c c c}
& New York & California & Texas & United States\\\hline
\# of Operands        & 13 & 13 & 14 & 14\\\hline  
\# of Buffers         & 7,686 & 16,754 & 19,7108 & 473,321\\\hline
\# of Capabilities    & 43,766 & 100,349 & 1,430,588 & 3,130,235\\\hline
\# of Elements in $M_\rho^+$. & 40,921 & 94,377 & 1,422,463 & 3,028,855\\\hline
\# of Elements in $M_\rho^-$. & 42,778 & 99,796 & 1,425,699 & 3,070,083\\\hline
HFIT Sparsity              & 4.6484e-5 & 2.2221e-5 & 1.4848e-6 & 6.7354e-7\\\hline
HFAM Sparsity              & 1.0618e-4 & 4.8221e-5 & 2.8641e-6 & 1.4259e-6\\\hline
Capability Land Density (capabilities/sqr mile) & 0.8022 & 0.6130 & 5.3262 & 1.0036\\\hline
Capability Population Density (capabilities/ppl) & 2.2104e-3 & 2.5599e-3 & 4.8495e-2 & 1.0206e-2\\\bottomrule
\end{tabular}
\end{center}
\vspace{-0.2in}
\end{table*}

As expected by the relative size of population and land area, Texas's energy infrastructure is the largest of the three states with California presenting the second largest and New York being the smallest in terms of the number of point-resources (buffers).  The same ordering holds true for the number of edges connecting the buffers in each of the states with TX presenting the most edges and NY presenting the least.  All three states are of course included in the full USA model including all of their buffers and edges.  There are 9,115 edges in NY, 20,674 edges in CA, 179,895 edges in TX, and 511,802 edges in the USA.  This corresponds to an adjacency matrix sparsity of 1.5438e-4, 7.3661e-5, 4.6306e-6, and 2.2848e-6 for NY, CA, TX, and the USA respectively.  As expected, the larger the system the more sparse the adjacency matrix becomes. While the three states display energy infrastructures of differing scales, they all show a diverse energy mixture that utilize the same 13 operands with Texas and the USA additionally utilizing solid biomass feedstock.  Also, despite these absolute measures, NY's population density is significantly higher than that of Texas, California, or the USA with TX and the USA having the lowest population densities.  This means that, if all else is held equal, New Yorkers receive more energy infrastructure benefits where Californians and Texans and the USA must spend more in energy infrastructure costs.

Table \ref{Ta:FGS} also presents the buffers per land and population densities.  When looking at the buffers per square mile density, CA had the smallest followed by NY, USA, then TX reporting 0.1023, 0.1409, 0.1518, and 0.7338 respectfully.  This interesting result shows the impact that TX has on influencing the USA's energy infrastructure as a whole. TX being such a large percentage of the USA's infrastructure, it brings up the buffer per square mile density above other states like NY or CA.  This is also seen in the buffers per population density with NY having the lowest followed by CA, USA, and TX respectfully.  These trends can also be tracked through the edges per square mile or population density.  Just as with buffers, CA has the lowest edges per square mile density followed by NY, USA, and TX respectfully.  Then when comparing the edges per population NY and CA switch, presenting NY with the lowest density followed by CA, USA, and TX respectfully.  These density measures can be used as a course indicator for the energy investment costs in each region. The lower the resource per population density, the greater the proxy cost per individual.  Similarly, the lower the resource per square mile density, the more spread out the cost is across the region's landscape.  

\subsection{Hetero-Functional Graph Statistics}\label{HFGStats}
The statistics of the hetero-functional graphs for NY, CA, TX, and the USA are presented in Table \ref{Ta:HFGS} similarly to those of the formal graphs.  Trends of sparsity in the HFITs match those of the formal graph adjacency matrices.  When comparing the number of capabilities in each model, NY has the least, followed by CA, TX, and the USA with $(43,766)$, $(100,349)$,  $(1,430,588)$, and $(3,130,235)$ capabilities respectively.  Unsurprisingly, the trend in sparsity of the HFITs follow that of the formal adjacency matrix.  NY's HFIT is twice as dense as CA's, which 31 times as dense as TX's, and which is 69 times as dense as the USA's HFIT.  Additionally, across all regions,  the negative HFIT has more filled elements than the positive indicating a greater amount of capabilities that pull operands from capabilities rather than injecting them.  A network topology with more capabilities that pull operands than inject operands is synonymous with a system that gathers and and refines operands than distributes and decomposes them. In the largest sense, this mathematical result is consistent with our understanding of the AMES which collects raw energy commodities (especially coal, oil, and natural gas) and delivers them as a relatively few ``high- grade" energy commodities like electricity.  The heterofunctional adjacency matrix can be calculated directly from the HFITs so as to link capabilities togther.  Again, as expected, these sparse matrices follow the same density trend as the HFIT with NY having the most dense matrix followed by CA, TX, and the USA with sparsity values of 1.0618e-4, 4.8221e-5, 2.8641e-6, and 1.4259e-6 respectfully.  

Table \ref{Ta:HFGS} also presents the capabilities per land and population densities.  When looking at the capabilities per square mile density, CA had the smallest followed by NY, USA, then TX reporting 0.6130, 0.8022, 1.0036, and 5.3262 respectively.  The ordering of capabilities per square miles density is the same trend as presented with buffers per square miles in the formal graph statistics.  Similarly, it shows the impact that states like TX have on influencing the full USA's energy infrastructure trends.  Additionally, due to the nature of allocating function onto form,  there are many more capabilities than buffers which brings the value of the density over land statistics up.  The increase in density values is also seen in the capabilities per population density with NY having the lowest followed by CA, USA, and TX respectfully.  These density measures can be used as an indicator for the benefit of each regions energy system.  The higher the capabilities per population density, the greater the amount of functionality (or benefits) from the infrastructure the population experiences.  Similarly, the higher the capability per square mile density, the more infrastructure benefits can be accessed across the regions landscape.  


\subsection{Formal Graph Resource Distribution}\label{FGResDist}
\begin{figure*}[t!]
\vspace{-0.1in}
\includegraphics[width=7.25in]{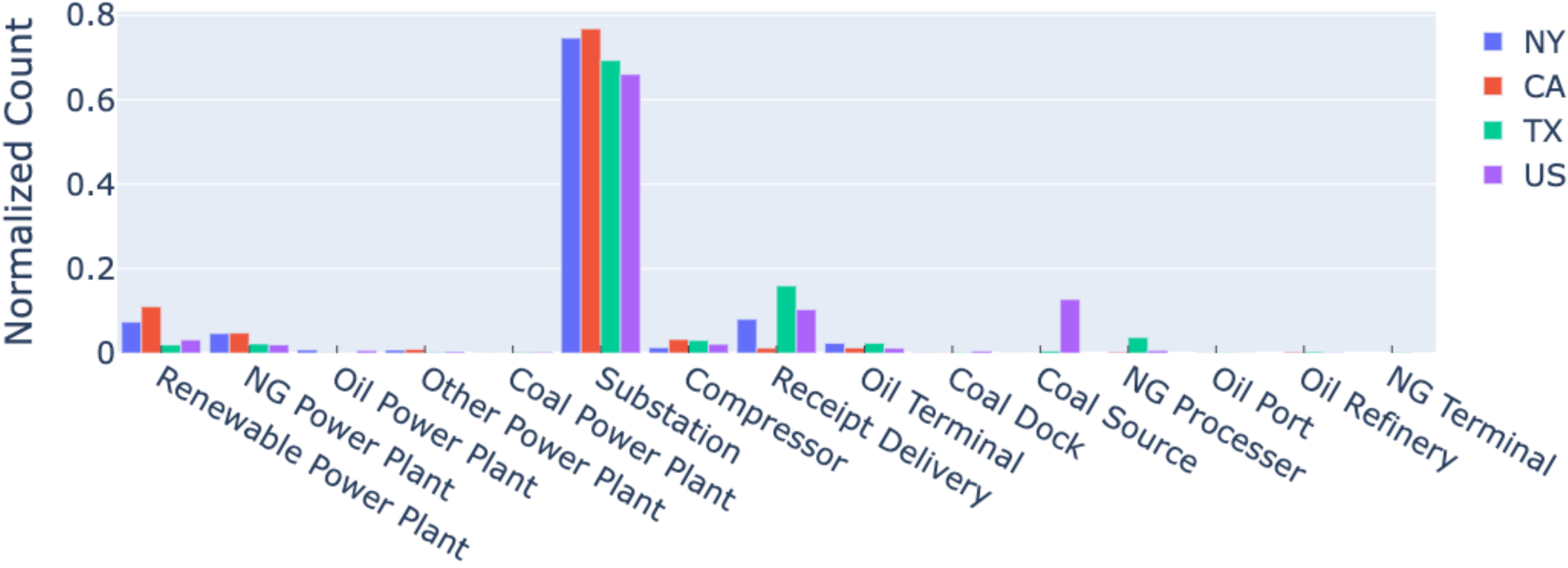} 
\caption{Normalized distribution of buffer types for NY, CA, TX, and the full USA.}
\vspace{-0.1in}
\label{Fig:nodeDist}
\end{figure*}

While it is important to assess the number of buffers (e.g. point energy facilities) in the multi-energy infrastructure of the three American states, it is also important to differentiate them by type.  Fig. \ref{Fig:nodeDist} shows that 74.6\%, 76.8\%, 69.3\%, and  66\% of the buffers in the formal graph are electric power substations in the states of NY, CA, TX, and the USA respectively.  The high percentage of substations reflects the highly ubiquitous nature of the electric power system in all four regions.  Furthermore, another 13.5\%, 16.8\%, 4.6\%, and 6.4\% of buffers are devoted towards electric power generation facilities (of various types) for NY, CA, TX, and the USA respectfully.  Because coal, oil, and natural gas are very dense approximate forms of energy their processing facilities for these types of energy have very strong economies of scale.  Therefore, there is a trend towards centralization and a small number of point facilities for energy conversion.  California, notably, has a greater shift toward the electric grid with a greater presence of substations and power plants than NY and TX.  Meanwhile, Texas is notably a large producer and trader of fossil fuels in the USA, and thus has the infrastructure to match.  That NY has a greater reliance on oil and gas facilities is likely a byproduct of it being located in a colder climate.  In opposition to the need for fossil fuels for space heating in cold environments, both TX and CA depend more on electricity for cooling spaces in their warmer climates and can make a greater use of renewable energy resources for such a cause.

\begin{figure*}[b!]
\vspace{-0.1in}
\includegraphics[width=7.25in]{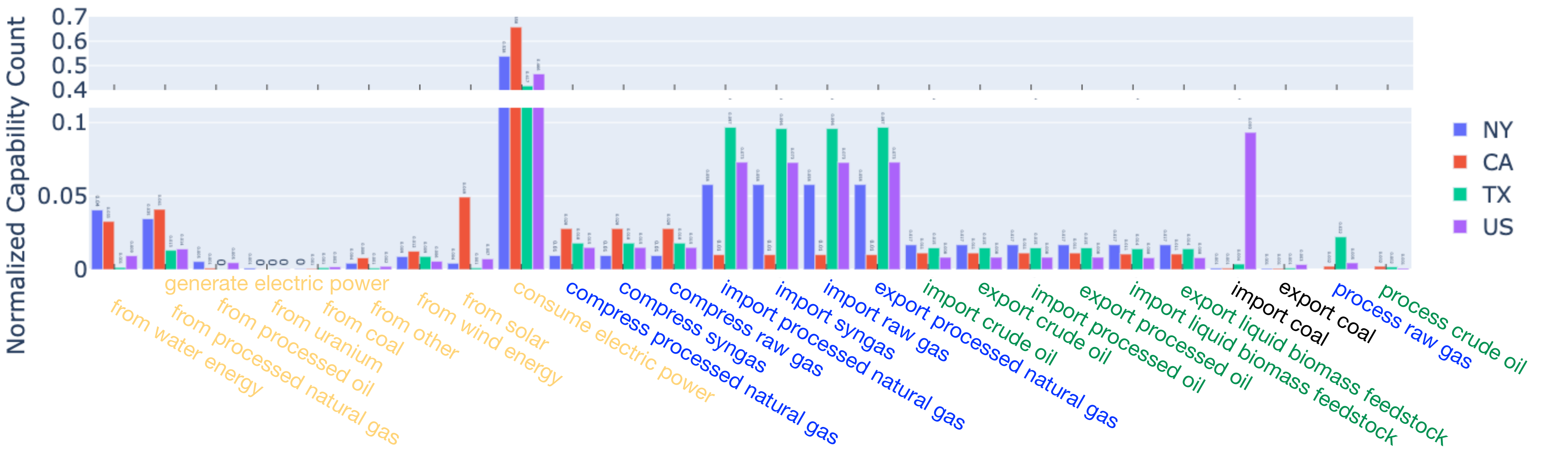} 
\caption{Normalized capability distribution of NY, CA, TX, and the USA.}
\vspace{-0.1in}
\label{Fig:CapabilityDist}
\end{figure*}
The first massive peak around substations reflects that there are more substations than any other type of node by far.  In addition to identifying the prominent dependence on electric power in American life, the peak also shoes that the transmission system comes a lot closer to the grid periphery of electric power systems than the corresponding distribution systems for coal, oil, or natural gas.  By taking a data driven approach on the Platts data, the discrepancies of how close each systems transmission level assets reach towards the system periphery becomes much more apparent.  For the electric grid, the system boundary ends at substations; ignoring the distributions system assets and consumption and generation at homes and busineses.  When looking at oil and natural gas, the system boundary stops at the terminals and ports where they are ultimately distributed outward.  By taking a data-driven approach on the Platts data, tankers and smaller gas lines delivering fuel for residential and industrial use are also ultimately omitted.  As a result, gas stations, homes, and other retail aspects of the oil and gas industries are not included.  Additionally, with coal being sold between commercial entities and not going out to individual consumers, the number of coal facilities needed to distribute to the demand is quite limited.  Despite the system boundary limitations of the Platts data, the analysis speaks to the state of the existing infrastructure and how much easier the electric power mode is to distribute than any other. The ease of electric power transportation, along with the potential reduction in carbon emissions, is one of the advantages to  electrifying energy demands as a part of the sustainable energy transition.

\begin{figure}[h!]
\vspace{-0.1in}
\includegraphics[width=3.5in]{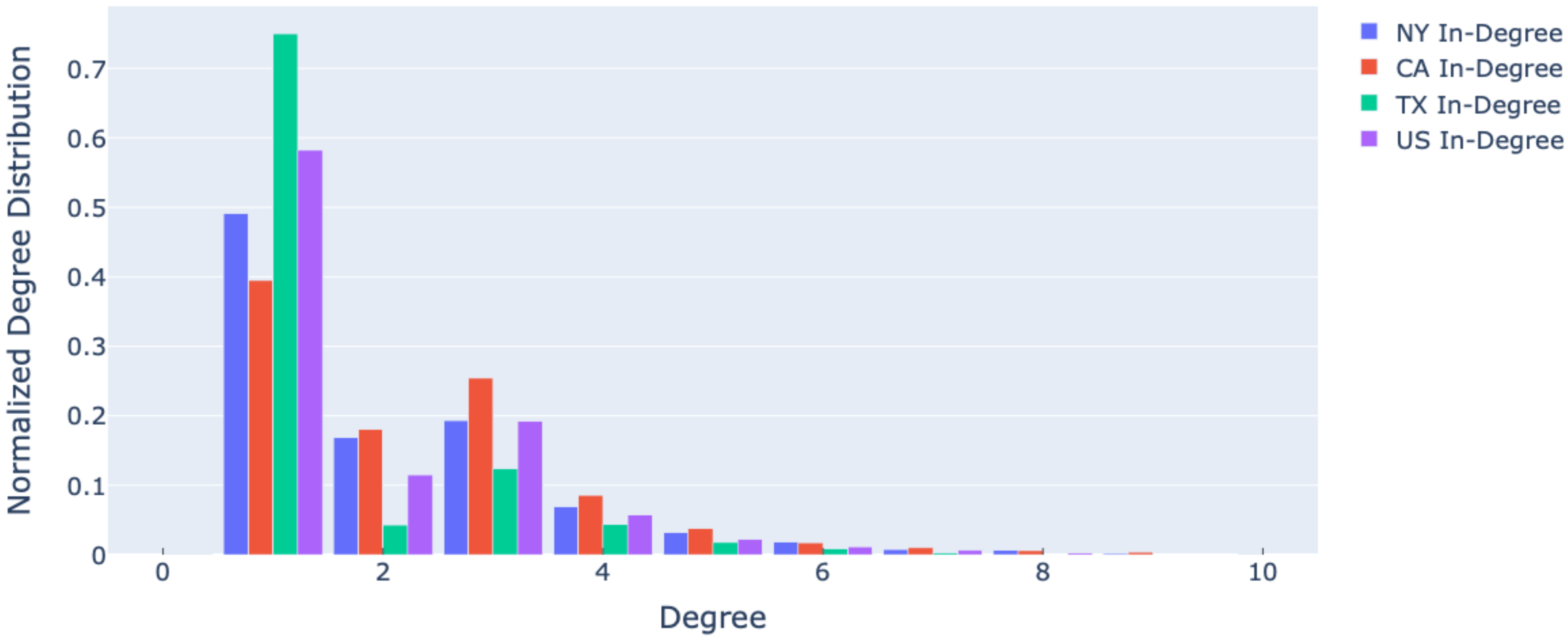} 
\caption{Normalized formal graph degree distribution of NY, CA, TX, and the USA.}
\vspace{-0.1in}
\label{Fig:degreeDist}
\end{figure}

Beyond the number and type of point-energy-facilities, the formal graph also measures their interconnectedness.  Despite the heterogeneity of point-energy-facilities and the sparsity of the four formal graphs of each state and the full USA, Figure \ref{Fig:degreeDist} shows that the formal graph degree distributions for the all four regions are remarkably similar.  In contrast to the well-known power-law degree distribution for electric power systems \cite{Amaral:2000:00,Albert:2004:00,Albert:2000:00,Barthelemy:2011:00,Barabasi:1999:00}, these multi-energy systems have degree distributions with a rather unusual shape.  Notability, all four regions peak at a single degree, then have a major lack of nodes with a degree 2.  After the initial dip, the degree distribution jumps back up to nodes more prevalent with a degree 3 before tailing off into an exponential decay.  Each energy subsystem likely contributes its own power law degree distribution so that the degree distributions depicted in Figure Figure \ref{Fig:degreeDist} are actually a composition of phenomena associated with the delivery of each energy commodity.   These differing contributing factors are further investigated through the hetero-functional graph distributions reported in the following sections. 

\subsection{Hetero-functional Capability Distributions}\label{HFGCapDist}

\begin{figure}[htb!]
\vspace{-0.1in}
\includegraphics[width=3.5in]{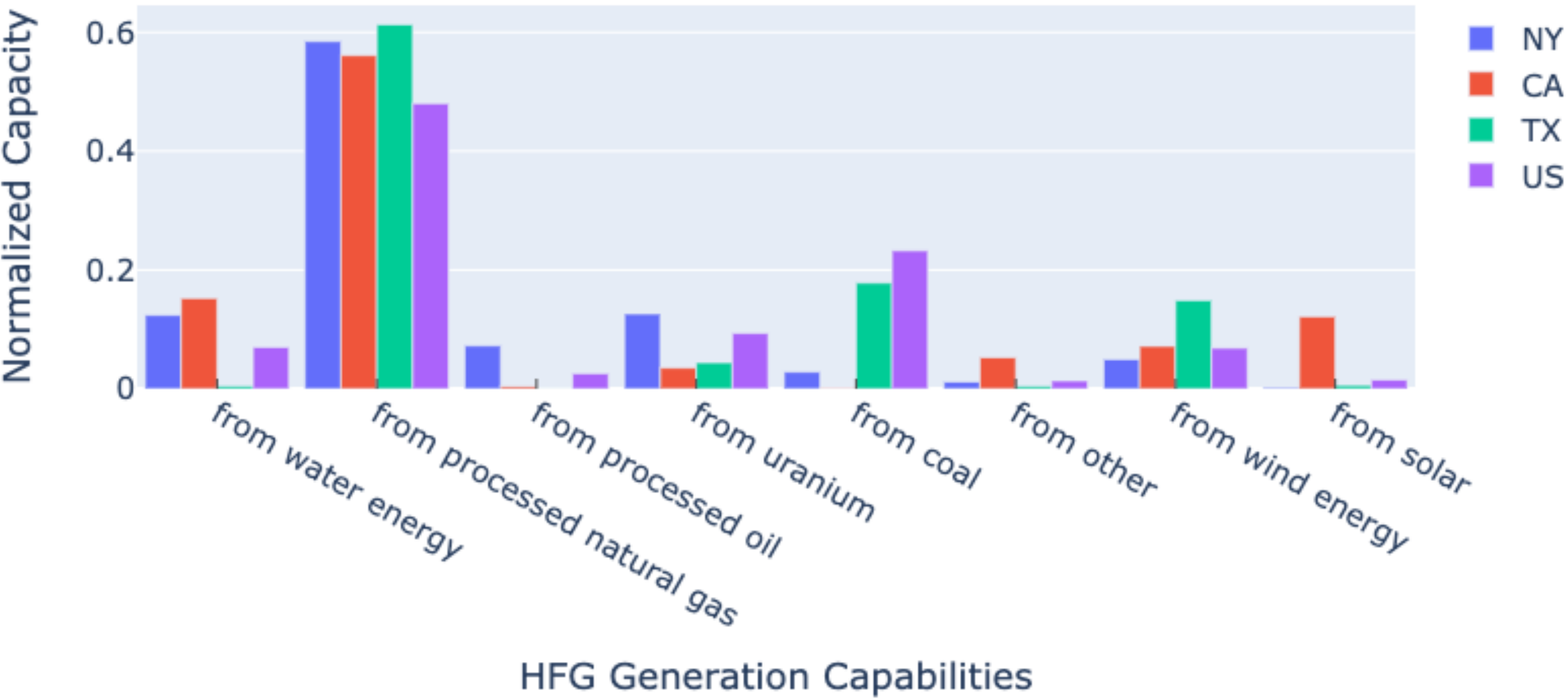} 
\caption{Normalized Electric Generation Capacity by fuel source for NY, CA, TX, and the full USA.}
\vspace{-0.1in}
\label{Fig:CapacityDist}
\end{figure}

In addition to looking at the type of nodes present in each region, the presence of different capabilities can also be compared.  As such, the normalized counts of the different capabilities are presented in Figure \ref{Fig:CapabilityDist} for NY, CA, TX, and the full USA.  Figure \ref{Fig:CapabilityDist} visualizes the prevalence of differing energy mixtures for each region as determined by their existing capabilities.  Figure \ref{Fig:CapabilityDist} shows a large spike in electric power capabilities similar to the formal graph node distributions shown in Figure \ref{Fig:nodeDist}.  Another notable trend is that TX and the full USA have very similar capability distributions; presumably by virtue of the large size of the TX energy system relative to the full USA.  The striking difference between TX and the USA is that the USA has an abundance of coal resources while TX has a very minimal amount.  The lack of coal in TX is likely a result of the coal sources residing most prominently in the Appalachian region. Additionally, NY, TX, and the USA have a heavy reliance on natural gas functionality.  All three regions have heavy spikes in functionality related to importing and exporting gas resources.  NY, as relatively cold northeastern state, relies heavily on natural gas for space heating during the cold winters.  In contrast, TX's natural gas functionality comes as a result of its large oil and gas economy.  The USA's capability distribution mixes these two dependencies.  Alternatively, CA is much more reliant on electric power generation as it has space conditioning requirements are largely dirven by electrified cooling rather natural-gas based heating. Notably, the functionality of solar, hydro and natural gas electric generation are most prominent as a result of CA's sustainable energy transition and the flexibility of natural gas electric generation to balance variable energy resources (VERs).  These natural gas electric generation capabilities provide valuable ramping functionality to balance the electric grid through sharp upward and downward ramps in response to VER generation and in demand levels.  These results are further confirmed and supported by the results in Figure \ref{Fig:CapacityDist} which presents the normalized capacities of each electric generation capability normalized by region.

As the sustainable energy transition requires an electrification of the demands placed on the AMES, it becomes important to investigate the electric power generation mix.  In accordance with single energy mode analysis, when the electric power generation capabilities in Figure \ref{Fig:CapabilityDist} are weighted by generation capacity, it produces Figure \ref{Fig:CapacityDist} which shows the electric generation mix for the three states and the USA.  The main source of generation is found to be natural gas for all four presented regions.  The high presence of natural gas generation indicates that the need for quick response generation sources to support VERs for NY, CA, TX, and the USA in addition to low fuel costs has lead to a large dependency on natural gas for electric power generation.  Additionally, NY sees a larger generation capacity from processed oil and nuclear power facilities than the other regions.  Again, the reliance on oil-fired generation stems from the need for ``peaker" units that support electrified heating in the coldest winter periods.CA's commitment to renewable energy generation is demonstrated in its solar and hydro capacity.  While Texas has a large investment in wind power.  Both TX and the USA also have significant capacity in coal-fired electric power generation.  In contrast, Figure \ref{Fig:CapacityDist} shows CA's and NY's energy transition to cleaner fuel sources and away from coal.  Finally, the electric power generation mix in Figure \ref{Fig:CapacityDist} demonstrates that when HFGT is restricted to a single commodity network (e.g electricity), it is able to reproduce the familiar analytics associated with individual (and often siloed) infrastructures.  

\begin{figure}
  \begin{subfigure}{3.5in}
    \centering\includegraphics[width=3.5in]{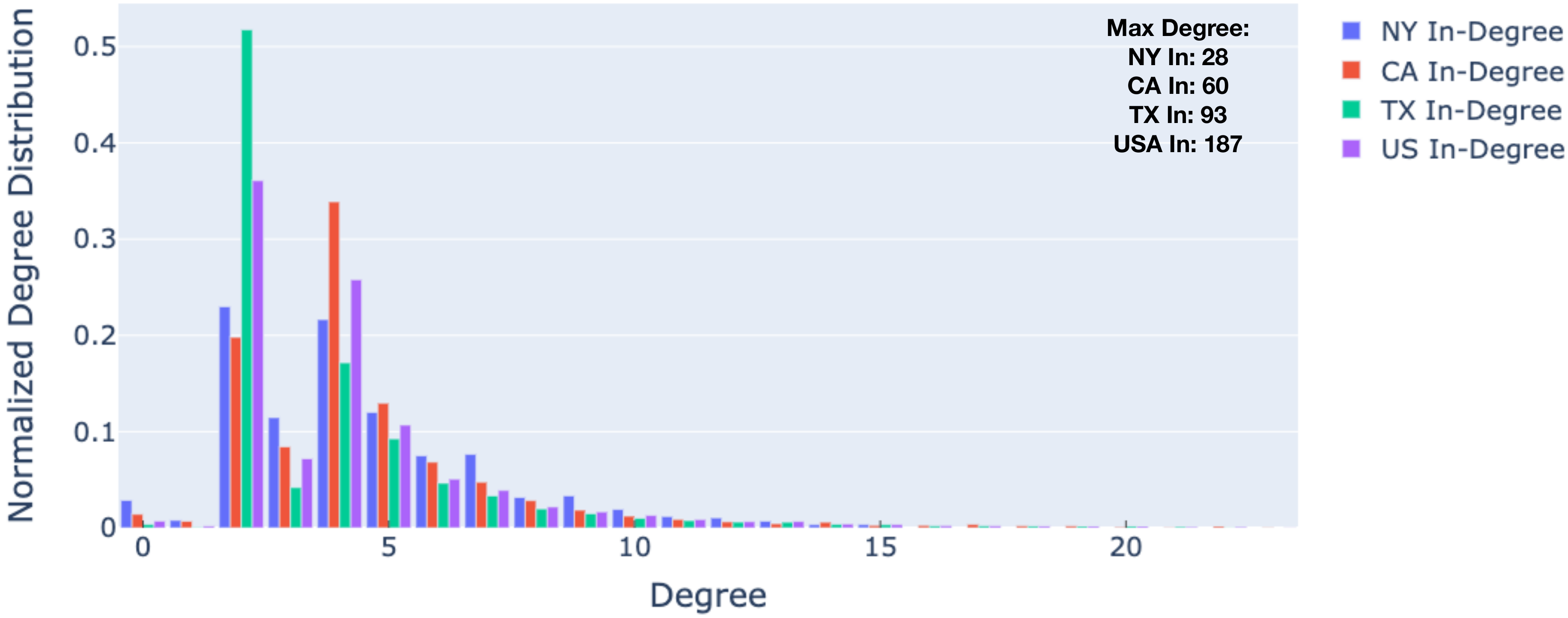}
    \caption{In-degree}
    \label{Fig:HFGInDegreeDist}
  \end{subfigure}
  \begin{subfigure}{3.5in}
    \centering\includegraphics[width=3.5in]{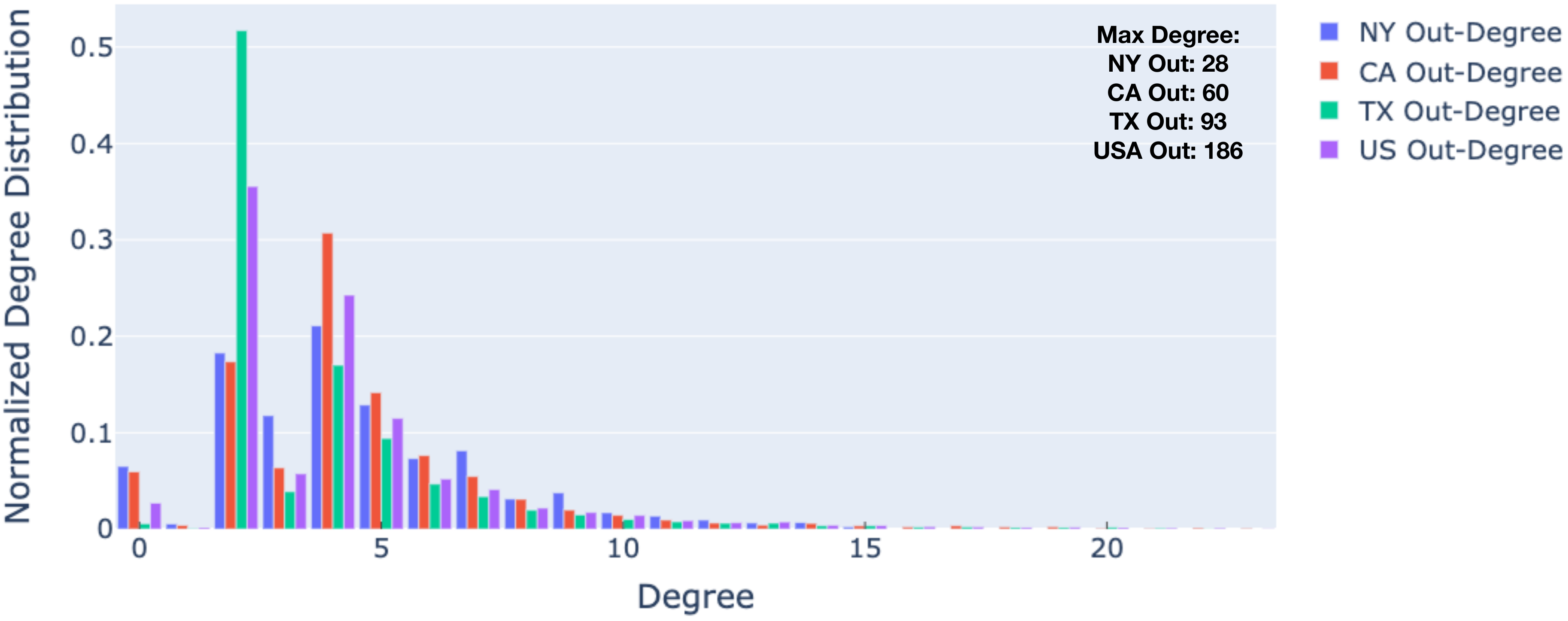}
    \caption{Out-degree}
    \label{Fig:HFGOutDegreeDist}
  \end{subfigure}
  \caption{Normalized hetero-functional graph degree distribution of NY, CA, TX, and the USA.}
  \label{Fig:HFGDegreeDist}
\end{figure}

Just as the degree distributions of the AMES FGs can be presented so can the degree distributions for the HFGs.  Figures \ref{Fig:HFGInDegreeDist} and \ref{Fig:HFGOutDegreeDist} show the AMES' HFG in-degree and out-degree distributions respectively.  As HFGs describe the logical sequence of capabilities, they are fundamentally directed in nature.  As such, both the in-degree and out-degree must be presented.  These plots present similar trends to those seen in the FG degree distributions shown in Figure \ref{Fig:degreeDist}.  Neither the FG degree distribution nor the HFG degree distributions follow a traditional power law decay, but the long tails in each plot suggest that there is some underlying power law behaviors.  From the HFG degree distributions, it is clear that there is some combination of exponentially decaying degree distributions from each of the four subsystems composing the AMES.  Interestingly, the general shape of both the in and out degree plots are very similar.  Both plots have normalized distributions that bottom out by degree 20 and have peaks at degrees 2 and 4.  Additionally, all four regions also have a major dip in their degree distributions at degree 3.  While the four regions all seem to have some underlying power law behaviors, each region presents their own combination of decays.  The differing degree distribution combinations suggest that there are geographical dependencies that influence the degree distributions.  Ultimately, while the results in Figure \ref{Fig:HFGDegreeDist} confirm those of Figure \ref{Fig:degreeDist}, the HFG model invites further investigation (in the following section) by classifying the HFG's capabilities by their underling process.

\subsection{Hetero-functional Graph Process Degree Distribution}\label{HFGProcDist}

\begin{figure*}[!t]
\vspace{-0.1in}
\centering
\includegraphics[width=6.5in]{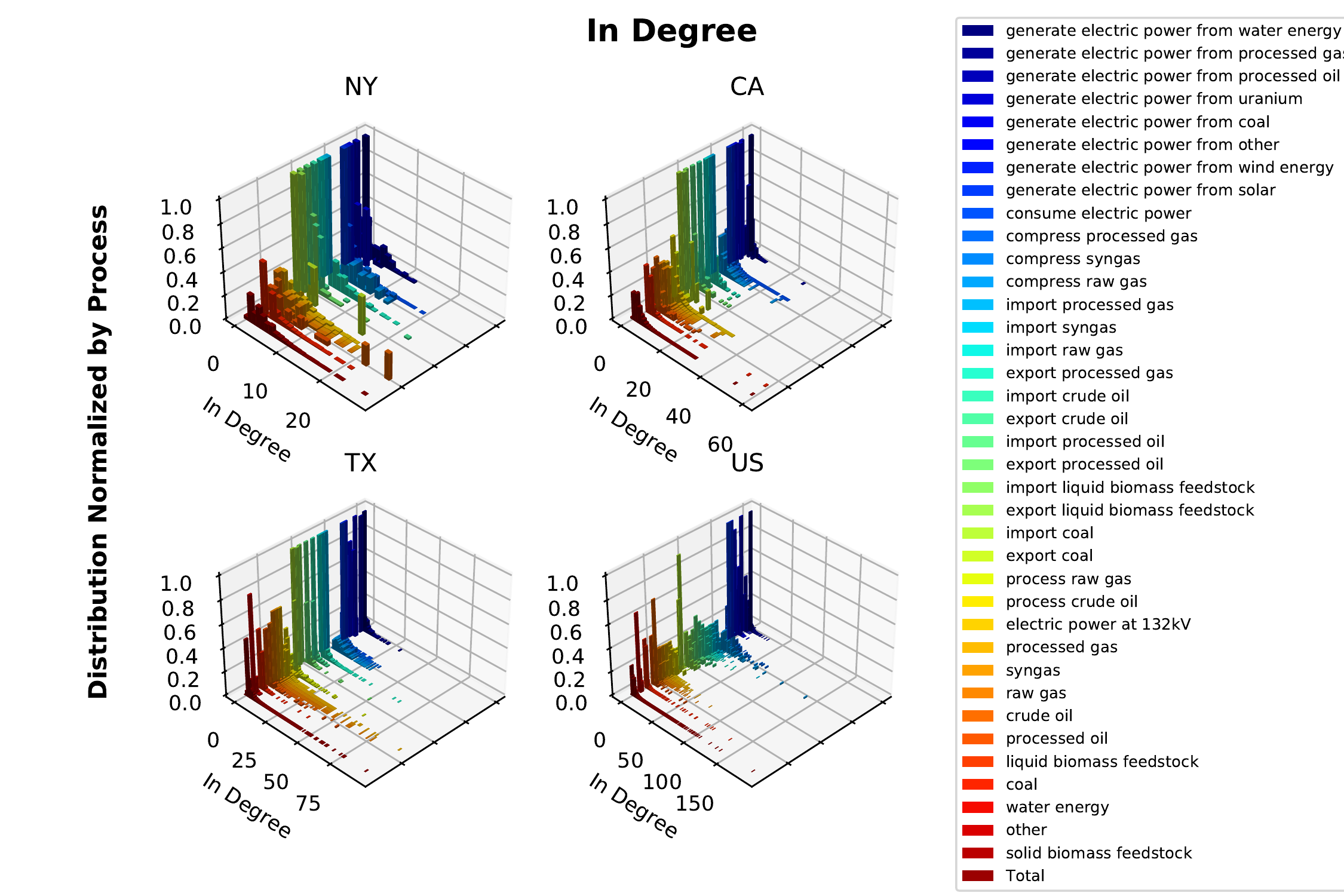}
\vspace{-0.1in}
\caption{In-Degree distribution of all Processes}
\vspace{-0.1in}
\label{Fig:processIn}
\end{figure*}

\begin{figure*}[!t]
\vspace{-0.1in}
\centering
\includegraphics[width=6.5in]{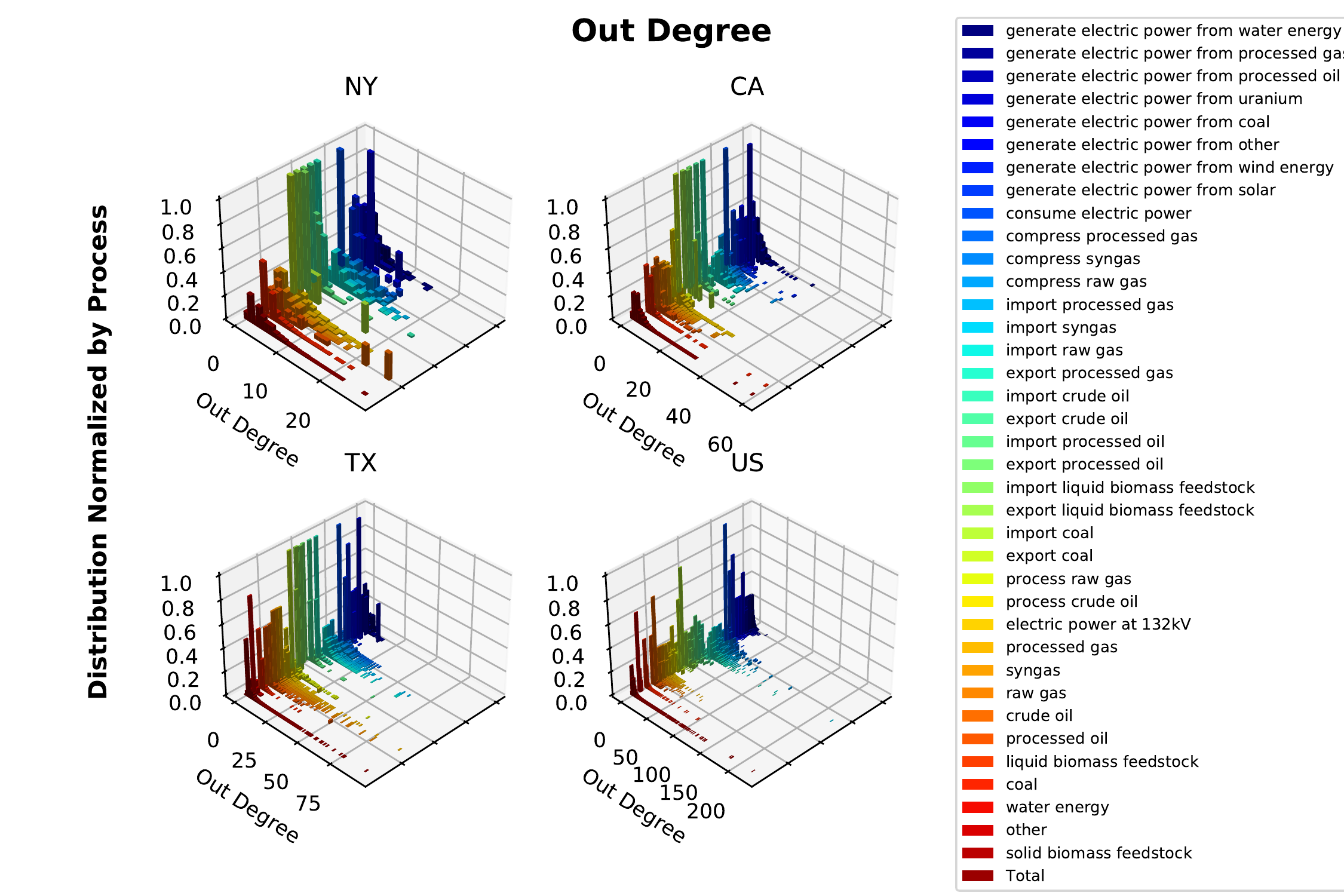}
\vspace{-0.1in}
\caption{Out-Degree distribution of all Processes}
\vspace{-0.1in}
\label{Fig:processOut}
\end{figure*}

Figures \ref{Fig:processIn} and \ref{Fig:processOut} present three dimensional (3D) degree distributions of the AMES' HFG where the third dimension classifies the node-capabilities by their underlying process.  This choice of classification is of critical analytical importance.  As discussed extensively in prior works\cite{Schoonenberg:2019:ISC-BK04,farid:2021:00}, capabilities can be associated with either point-type buffers or line-type transportation resources.  Therefore, a classification scheme based on buffers is both incomplete and logically inconsistent with the nature of a HFG.  In the meantime, a classification schmee based on the capabilities' (input or output) operands runs the risk of double counting the capabilities because each capability can have more than one type of operand (e.g. a natural gas electric power plant).  Therefore, the most straightforward way of classifying capabilities into mutually-exclusive and totally exhaustive sets while preserving the objectivity of the underlying statistics is based upon the capabilities' underlying process.

Returning to Figures \ref{Fig:processIn} and \ref{Fig:processOut}, the former presents the 3D in-degree distribution while the latter presents the 3D out-degree distribution.  Both figures reveal, for the first time, the underlying nature of the AMES' hetero-functional structure.  With rare exception, the AMES exhibits a power-law degree distribution for each set of capabilities classified by process.  These figures provide a fascinating empirical result.  Previous works demonstrate that HFGs of single operand networks have power law degree distributions just like formal graphs of single operand networks\cite{Thompson:2020:01,Thompson:2021:00}.  Now, this paper shows that a HFG of a multi-operand network exhibits a power law degree distribution for each type of process so that the final HFG degree distribution in Figure  \ref{Fig:HFGDegreeDist} shows a superposition of the degree distributions associated with each type of process.  In contrast, such an empirical result can not be reached using a formal graph with buffer-nodes because of the information loss caused by the sums in Eq. \ref{Eq:FG1} and \ref{Eq:FG2}.  In other words, HFGT has confirmed well-known results in the network science literature and successfully generalized them for systems that are fundamentally hetero-functional in nature.  

This theoretical insight has direct practical relevance to the fundamental systems science principles underlying the sustainable energy transition.  The 3D degree distributions in Figures \ref{Fig:processIn} and \ref{Fig:processOut} show that the AMES' structure is evolving at different rates depending on the underlying process of each capability.  In other words, from a graph theory perspective, the sustainable energy transition can be understood as \emph{network phenomena} where the decommissioning of carbon-intensive processes (e.g. oil and gas refining and coal-fired electric power generation) is occurring at a certain rate, and carbon-light processes (e.g. wind and solar power generation) is occurring at a different rate.  Similarly, broad trends toward electrification (or fuel switching) is a superposition of the network phenomena driving the consumption of carbon-intensive fuels relative to consuming electricity.  Said differently, HFGT points to the network phenomena underpinning ``infrastructure lock-in" like homeowners who wish switch to heat-pump technology but find themselves ``locked in" to furnace-based heating.  Similarly, HFGT points to the network phenomena underpinning ``infrastructure platforms" like car owners who wish to buy an electric vehicle but find themselves worrying about local charging infrastructure.  Ultimately, a deeper study of the power laws in Figures \ref{Fig:processIn} and \ref{Fig:processOut} is warranted because the relative rates underpin the success of the sustainable energy transition.  From a policy perspective, the sustainable energy transition requires that the rate of integrating certain types of sustainable energy capabilities be ``sped-up" relative to more carbon intensive ones.  

\section{Conclusions and Policy Implications}\label{Conclusions}

This paper uses a data-driven, MBSE-guided approach to develop open-source software that produces open structural models of the American Multi-modal Energy System.  It is part of a larger NSF project entitled ``American Multi-Modal Energy System Synthetic \& Simulated Data (AMES-3D)" which seeks to produce open-source structural and behavioral models of the American Multi-modal Energy System.  The creation of open-source software and open-data models of the AMES fills an important need in open, citizen-based science in America's sustainable energy transition.  It also provides one of the few multi-energy system datasets in which to advance fundamental methods.  The AMES' structural models are inferred from the Platts Map Data Pro GIS dataset and is complemented by the previously developed American Multi-modal Energy System Reference Architecture \cite{Thompson:2020:00}.  Together, these two data sources serve as the basis for an XML-based input data file for the open-source hetero-functional graph theory toolbox.  

This paper specifically reports the hetero-functional incidence tensor, the formal graph adjacency matrix and hetero-functional graph adjacency matrix statistics for the multi-energy infrastructure systems for the states of NY, CA, TX, and the full USA. Here, the application of hetero-functional graph theory facilitates a nuanced analysis that respects the heterogeneity in this highly interdependent system-of-systems.  The paper finds that the geography and sustainable energy policies of the states are deeply reflected in the structure of their multi-energy infrastructure.  Because New York's cold north eastern climate drives heating demand, it has a multi-energy system with a greater emphasis on oil and gas.  In the meantime, California's warm climate is reflected in a multi-energy system with a greater emphasis on electric power systems.  Additionally, Texas has a large oil and gas economy and thus has a large percentage of energy infrastructure pertaining to these fossil fuels but also has build out of a tremendous amount of wind energy for electric generation.  Along these lines, California and Texas have also geared their natural gas resources infrastructure towards electric power generation to support their growing reliance on variable energy resources.  These trends appear as components of the AMES as a whole.  Additionally, states with a large energy infrastructure like Texas have a greater impact on shaping the USA's energy infrastructure.  It is also important to note the import and export functionality of these fuel sources are also very prominent.  Identifying the abundance of import and export functionality is important for the evolution of a multi-modal energy system for two reasons.  Import and export functionality provides open interfaces to new modes of energy delivery such as hydrogen.  Important export functionality in a HFG also creates a theoretical foundation from which to investigate energy systems with relatively large import/export economies (e.g. Australia).  Finally, through the analysis of HFG degree distributions and their associated processes,  it becomes clear that power-law degree distributions are fundamentally tied to the processes rather than point-resources (i.e. buffers).  

From the perspective of understanding the dynamics of the sustainable energy transition, the power laws imply that for each type of process, the "popular" (e.g. most well-connected) become even more so.   The degree distribution power law for each process reinforces the connections of the capabilities with more connected capabilities.  The AMES' HFG has a power law associated with every type of  process and each power law reflects a rate of evolution or adoption. 

Consequently, many new sustainable energy process technologies may require policies, at least initially, that support them relative to incumbent process technologies.  These initial policies are likely to instigate network-driven positive feedback loops that accelerate adoption rates.  For example, as renewable energy sources are adopted, it expands the electric grid itself, which in turn supports electrification on the AMES' demand side.   These implications are also applicable to hydrogen as it becomes a more viable and prominent technology.  As this work was driven by the Platts GIS data, it was ultimately confined to incumbent conventional energy sources, rather than new energy pathways tied to hydrogen, synthetic fuels and bio-energy.  Naturally, the transparent natures of the AMES reference architecture lends itself to revision so as to include these new energy carriers based upon first-principle process physics.  Finally, the power law results found in Figures \ref{Fig:processIn} and \ref{Fig:processOut} present a ripe opportunity for further investigation. 

Beyond the structural analysis presented here, the AMES' structural model presents multiple avenues for future open-science research.  First, behavioral data can be incorporated so as to develop a physically-informed machine- learning behavioral model of the AMES.  Secondly, the AMES can be studied rigorously for its sustainability and resilience properties using novel methods rooted in hetero-functional graph theory.  Finally, scenarios varying the energy mixtures seen in these regions and across the full USA can be investigated to explore potential sustainable energy transitions.  

\vspace{-0.1in}
\section{Acknowledgement}
This work is supported by funding from the National Science Foundation.  
\vspace{-0.1in}

\bibliographystyle{IEEEtran}
\bibliography{AMES_Inst_Lib.bib}

\begin{thebibliography}{10}
\providecommand{\url}[1]{#1}
\csname url@samestyle\endcsname
\providecommand{\newblock}{\relax}
\providecommand{\bibinfo}[2]{#2}
\providecommand{\BIBentrySTDinterwordspacing}{\spaceskip=0pt\relax}
\providecommand{\BIBentryALTinterwordstretchfactor}{4}
\providecommand{\BIBentryALTinterwordspacing}{\spaceskip=\fontdimen2\font plus
\BIBentryALTinterwordstretchfactor\fontdimen3\font minus
  \fontdimen4\font\relax}
\providecommand{\BIBforeignlanguage}[2]{{%
\expandafter\ifx\csname l@#1\endcsname\relax
\typeout{** WARNING: IEEEtran.bst: No hyphenation pattern has been}%
\typeout{** loaded for the language `#1'. Using the pattern for}%
\typeout{** the default language instead.}%
\else
\language=\csname l@#1\endcsname
\fi
#2}}
\providecommand{\BIBdecl}{\relax}
\BIBdecl

\bibitem{Moniz:2022:00}
\BIBentryALTinterwordspacing
E.~J. Moniz. (2022, July) A way forward on the climate crisis energy
  insecurity. [Online]. Available:
  \url{https://www.bostonglobe.com/2022/07/18/opinion/way-forward-climate-crisis-energy-insecurity/}
\BIBentrySTDinterwordspacing

\bibitem{action:2010:00}
C.~Action, ``Framework convention on climate change,'' 2010.

\bibitem{wec:2016:00}
WEC, ``World energy trilemma 2016: Defining measures to accelerate the energy
  transition,'' 2016.

\bibitem{bridge:2013:00}
G.~Bridge, S.~Bouzarovski, M.~Bradshaw, and N.~Eyre, ``Geographies of energy
  transition: Space, place and the low-carbon economy,'' \emph{Energy policy},
  vol.~53, pp. 331--340, 2013.

\bibitem{Vespignani:2010:00}
\BIBentryALTinterwordspacing
A.~Vespignani, ``The fragility of interdependency,'' \emph{Nature}, vol. 464,
  no. 7291, pp. 984--985, 2010. [Online]. Available:
  \url{https://doi.org/10.1038/464984a}
\BIBentrySTDinterwordspacing

\bibitem{Albert:2004:00}
R.~Albert, I.~Albert, and G.~L. Nakarado, ``Structural vulnerability of the
  north american power grid,'' \emph{Physical review E}, vol.~69, no.~2, p.
  025103, 2004.

\bibitem{Dong:2015:00}
G.~Dong, R.~Du, L.~Tian, and R.~Liu, ``Robustness of network of networks with
  interdependent and interconnected links,'' \emph{Physica A: Statistical
  Mechanics and its Applications}, vol. 424, pp. 11--18, 2015.

\bibitem{Jean-Baptiste:2003:00}
P.~Jean-Baptiste and R.~Ducroux, ``Energy policy and climate change,''
  \emph{Energy Policy}, vol.~31, no.~2, pp. 155--166, Jan 2003.

\bibitem{Kriegler:2018:00}
E.~Kriegler, G.~Luderer, N.~Bauer, L.~Baumstark, S.~Fujimori, A.~Popp,
  J.~Rogelj, J.~Strefler, and D.~P. Van~Vuuren, ``Pathways limiting warming to
  1.5c: a tale of turning around in no time?'' \emph{Philosophical Transactions
  of the Royal Society A: Mathematical, Physical and Engineering Sciences},
  vol. 376, no. 2119, p. 20160457, 2018.

\bibitem{Robert-Lempert:2019:00}
{Robert Lempert, Benjamin L. Preston, Jae Edmonds, Leon Clarke, Tom Wild,
  Matthew Binsted, Elliot Diringer, Brad Townsend}, ``Pathways to 2050
  alternative scenarios for decarbonizing the u.s. economy,'' Center for
  Climate and Energy Solutions, Climate Innovation 2050, 2019.

\bibitem{Rogers:2013:00}
J.~Rogers, K.~Averyt, S.~Clemmer, M.~Davis, F.~Flores-Lopez, D.~Kenney,
  J.~Macknick, N.~Madden, J.~Meldrum, S.~Sattler, and E.~Spanger-Siegfried,
  ``{Water-Smart Power: Strengthening the U.S. Electricity System in a Warming
  World},'' Union for Concerned Scientists, Cambridge, MA, Tech. Rep., 2013.

\bibitem{Lara:2020:00}
\BIBentryALTinterwordspacing
P.~Lara, M.~S{\'a}nchez, and J.~Villalobos, ``Enterprise modeling and
  operational technologies (ot) application in the oil and gas industry,''
  \emph{Journal of Industrial Information Integration}, vol.~19, p. 100160,
  2020. [Online]. Available:
  \url{https://www.sciencedirect.com/science/article/pii/S2452414X20300352}
\BIBentrySTDinterwordspacing

\bibitem{Panteli:2019:00}
M.~Panteli, E.~A.~M. Cese{\~n}a, R.~Moreno, and P.~Mancarella, ``Infrastructure
  planning under uncertainty: flexibility, resilience and multi-energy systems
  application tutorial,'' 2019.

\bibitem{Mejia-Giraldo:2012:00}
\BIBentryALTinterwordspacing
D.~Mejia-Giraldo, J.~Villarreal-Marimon, Y.~Gu, Y.~He, Z.~Duan, and L.~Wang,
  ``{Sustainability and resiliency measures for long-term investment planning
  in integrated energy and transportation infrastructures},'' \emph{Journal of
  Energy Engineering}, vol. 138, no.~2, pp. 87--94, Jun. 2012. [Online].
  Available: \url{http://dx.doi.org/10.1061/(ASCE)EY.1943-7897.0000067}
\BIBentrySTDinterwordspacing

\bibitem{masters:2013:00}
G.~M. Masters, \emph{Renewable and efficient electric power systems}.\hskip 1em
  plus 0.5em minus 0.4em\relax John Wiley \& Sons, 2013.

\bibitem{mokhatab:2012:00}
S.~Mokhatab and W.~A. Poe, \emph{Handbook of natural gas transmission and
  processing}.\hskip 1em plus 0.5em minus 0.4em\relax Gulf professional
  publishing, 2012.

\bibitem{Lurie:2009:00}
M.~V. Lurie, \emph{Modeling of Oil Product and Gas Pipeline
  Transportation}.\hskip 1em plus 0.5em minus 0.4em\relax Wiley-VCH Verlag GmbH
  \& Co. KGaA, 2009.

\bibitem{EIA:2014:11}
EIA, ``Coal market module of the national energy modeling system: Model
  documentation 2014,'' \emph{Independent Statistics and Analysis, U.S. Energy
  Information Administration, Department of Energy}, 2014.

\bibitem{Priyanka:2021:00}
\BIBentryALTinterwordspacing
E.~Priyanka, S.~Thangavel, X.-Z. Gao, and N.~Sivakumar, ``Digital twin for oil
  pipeline risk estimation using prognostic and machine learning techniques,''
  \emph{Journal of Industrial Information Integration}, p. 100272, 2021.
  [Online]. Available:
  \url{https://www.sciencedirect.com/science/article/pii/S2452414X21000704}
\BIBentrySTDinterwordspacing

\bibitem{EIA:2020:00}
EIA, ``Annual energy outlook 2020,'' \emph{Independent Statistics and Analysis,
  U.S. Energy Information Administration, Department of Energy}, 2020.

\bibitem{EIA:2017:01}
\BIBentryALTinterwordspacing
------, ``Availability of the national energy modeling system (nems) archive,''
  United States Energy Information Administration, Tech. Rep., 2017. [Online].
  Available: \url{https://www.eia.gov/outlooks/aeo/info_nems_archive.cfm}
\BIBentrySTDinterwordspacing

\bibitem{Johnson:2017:00}
J.~K. Barry W.~Johnson, ``Dear colleague letter: Simulated andsynthetic data
  for infrustructure modeling,'' March 2017.

\bibitem{Dori:2015:00}
D.~Dori, \emph{Model-based systems engineering with OPM and SysML}.\hskip 1em
  plus 0.5em minus 0.4em\relax Springer, 2015.

\bibitem{Friedenthal:2011:00}
S.~Friedenthal, A.~Moore, and R.~Steiner, \emph{{A Practical Guide to SysML:
  The Systems Modeling Language}}, 2nd~ed.\hskip 1em plus 0.5em minus
  0.4em\relax Burlington, MA: Morgan Kaufmann, 2011.

\bibitem{Thompson:2020:00}
D.~J. Thompson and A.~M. Farid, ``A reference architecture for the american
  multi-modal energy system,'' \emph{arXiv preprint arXiv:2012.14486}, 2020.

\bibitem{PlazasNino:2022:00}
\BIBentryALTinterwordspacing
F.~Plazas-Ni{\~n}o, N.~Ortiz-Pimiento, and E.~Montes-P{\'a}ez, ``National
  energy system optimization modelling for decarbonization pathways analysis: A
  systematic literature review,'' \emph{Renewable and Sustainable Energy
  Reviews}, vol. 162, p. 112406, 2022. [Online]. Available:
  \url{https://www.sciencedirect.com/science/article/pii/S1364032122003148}
\BIBentrySTDinterwordspacing

\bibitem{Millot:2020:00}
\BIBentryALTinterwordspacing
A.~Millot, A.~Krook-Riekkola, and N.~Ma{\"\i}zi, ``Guiding the future energy
  transition to net-zero emissions: Lessons from exploring the differences
  between france and sweden,'' \emph{Energy Policy}, vol. 139, p. 111358, 2020.
  [Online]. Available:
  \url{https://www.sciencedirect.com/science/article/pii/S0301421520301154}
\BIBentrySTDinterwordspacing

\bibitem{Dedinec:2016:00}
\BIBentryALTinterwordspacing
A.~Dedinec, V.~Taseska-Gjorgievska, N.~Markovska, T.~{Obradovic Grncarovska},
  N.~Duic, J.~Pop-Jordanov, and R.~Taleski, ``Towards post-2020 climate change
  regime: Analyses of various mitigation scenarios and contributions for
  macedonia,'' \emph{Energy}, vol.~94, pp. 124--137, 2016. [Online]. Available:
  \url{https://www.sciencedirect.com/science/article/pii/S036054421501453X}
\BIBentrySTDinterwordspacing

\bibitem{Yangka:2016:00}
\BIBentryALTinterwordspacing
D.~Yangka and M.~Diesendorf, ``Modeling the benefits of electric cooking in
  bhutan: A long term perspective,'' \emph{Renewable and Sustainable Energy
  Reviews}, vol.~59, pp. 494--503, 2016. [Online]. Available:
  \url{https://www.sciencedirect.com/science/article/pii/S1364032115016482}
\BIBentrySTDinterwordspacing

\bibitem{Thompson:2020:01}
D.~J. Thompson, W.~C. Schoonenberg, and A.~M. Farid, ``A hetero-functional
  graph analysis of electric power system structural resilience,'' in
  \emph{2020 IEEE Power Energy Society Innovative Smart Grid Technologies
  Conference (ISGT)}, 2020, pp. 1--5.

\bibitem{Thompson:2021:00}
------, ``A hetero-functional graph resilience analysis of the future american
  electric power system,'' \emph{IEEE Access}, vol.~9, pp. 68\,837--68\,848,
  2021.

\bibitem{Platts:2017:00}
\BIBentryALTinterwordspacing
Platts, ``Platts energy map data pro,'' {S\&P Global Platts}, Tech. Rep., 2017.
  [Online]. Available:
  \url{https://www.spglobal.com/platts/en/products-services/oil/map-data-pro}
\BIBentrySTDinterwordspacing

\bibitem{EIA:2021:01}
EIA, ``California state energy profile,'' \emph{EIA U.S. States Overview},
  February 2021.

\bibitem{EIA:2021:00}
------, ``New york state energy profile,'' \emph{EIA U.S. States Overview},
  October 2021.

\bibitem{EIA:2022:00}
------, ``Texas state energy profile,'' \emph{EIA U.S. States Overview}, May
  2022.

\bibitem{Cloutier:2010:00}
R.~Cloutier, G.~Muller, D.~Verma, R.~Nilchiani, E.~Hole, and M.~Bone, ``The
  concept of reference architectures,'' \emph{Systems Engineering}, vol.~13,
  no.~1, pp. 14--27, 2010.

\bibitem{Oberle:2019:00}
\BIBentryALTinterwordspacing
S.~Oberle and R.~Elsland, ``Are open access models able to assess today's
  energy scenarios?'' \emph{Energy Strategy Reviews}, vol.~26, p. 100396, 2019.
  [Online]. Available:
  \url{https://www.sciencedirect.com/science/article/pii/S2211467X19300896}
\BIBentrySTDinterwordspacing

\bibitem{Lopion:2018:00}
\BIBentryALTinterwordspacing
P.~Lopion, P.~Markewitz, M.~Robinius, and D.~Stolten, ``A review of current
  challenges and trends in energy systems modeling,'' \emph{Renewable and
  Sustainable Energy Reviews}, vol.~96, pp. 156--166, 2018. [Online].
  Available:
  \url{https://www.sciencedirect.com/science/article/pii/S1364032118305537}
\BIBentrySTDinterwordspacing

\bibitem{HOWELLS:2011:00}
\BIBentryALTinterwordspacing
M.~Howells, H.~Rogner, N.~Strachan, C.~Heaps, H.~Huntington, S.~Kypreos,
  A.~Hughes, S.~Silveira, J.~DeCarolis, M.~Bazillian, and A.~Roehrl,
  ``Osemosys: The open source energy modeling system: An introduction to its
  ethos, structure and development,'' \emph{Energy Policy}, vol.~39, no.~10,
  pp. 5850--5870, 2011, sustainability of biofuels. [Online]. Available:
  \url{https://www.sciencedirect.com/science/article/pii/S0301421511004897}
\BIBentrySTDinterwordspacing

\bibitem{EIA:2019:00}
EIA, ``The national energy modeling system: An overview 2018,''
  \emph{Independent Statistics and Analysis, U.S. Energy Information
  Administration, Department of Energy}, April 2019.

\bibitem{E3MLAB:2018:00}
E3MLab, ``Primes model version 2018 detailed model description,''
  \emph{E3MLAB}, 2018.

\bibitem{Muzhikyan:2018:SPG-W07}
\BIBentryALTinterwordspacing
A.~Muzhikyan, S.~O. Muhanji, G.~Moynihan, D.~Thompson, Z.~Berzolla, and A.~M.
  Farid, ``{ISO New England System Operational Analysis and Renewable Energy
  Integration Study: Final Project Report},'' Laboratory for Intelligent
  Integrated Networks of Engineering Systems. Thayer School of Engineering at
  Dartmouth., Hanover, NH, USA, Tech. Rep., 2018. [Online]. Available:
  \url{http://engineering.dartmouth.edu/liines/resources/TechReports/SPG-W07.pdf}
\BIBentrySTDinterwordspacing

\bibitem{Muhanji:2020:EWN-J44}
\BIBentryALTinterwordspacing
S.~O. Muhanji, C.~Barrows, J.~Macknick, and A.~M. Farid, ``{An Enterprise
  Control Assessment Case Study of the Energy-Water Nexus for the ISO New
  England System},'' \emph{Renewable and Sustainable Energy Reviews}, vol. 141,
  pp. 110\,766--110\,785, 2021. [Online]. Available:
  \url{https://doi.org/10.1016/j.rser.2021.110766}
\BIBentrySTDinterwordspacing

\bibitem{Stralen:2021:00}
\BIBentryALTinterwordspacing
J.~N.~P. van Stralen, F.~Dalla~Longa, B.~W. Dani{\"e}ls, K.~E.~L. Smekens, and
  B.~van~der Zwaan, ``Opera: a new high-resolution energy system model for
  sector integration research,'' \emph{Environmental Modeling \& Assessment},
  vol.~26, no.~6, pp. 873--889, 2021. [Online]. Available:
  \url{https://doi.org/10.1007/s10666-020-09741-7}
\BIBentrySTDinterwordspacing

\bibitem{Limpens:2019:00}
\BIBentryALTinterwordspacing
G.~Limpens, S.~Moret, H.~Jeanmart, and F.~Mar{\'e}chal, ``Energyscope td: A
  novel open-source model for regional energy systems,'' \emph{Applied Energy},
  vol. 255, p. 113729, 2019. [Online]. Available:
  \url{https://www.sciencedirect.com/science/article/pii/S0306261919314163}
\BIBentrySTDinterwordspacing

\bibitem{Heendeniya:2020:00}
\BIBentryALTinterwordspacing
C.~B. Heendeniya, A.~Sumper, and U.~Eicker, ``The multi-energy system
  co-planning of nearly zero-energy districts -- status-quo and future research
  potential,'' \emph{Applied Energy}, vol. 267, p. 114953, 2020. [Online].
  Available:
  \url{https://www.sciencedirect.com/science/article/pii/S0306261920304657}
\BIBentrySTDinterwordspacing

\bibitem{Schoonenberg:2019:ISC-BK04}
\BIBentryALTinterwordspacing
W.~C. Schoonenberg, I.~S. Khayal, and A.~M. Farid, \emph{{A Hetero-functional
  Graph Theory for Modeling Interdependent Smart City Infrastructure}}.\hskip
  1em plus 0.5em minus 0.4em\relax Berlin, Heidelberg: Springer, 2019.
  [Online]. Available: \url{http://dx.doi.org/10.1007/978-3-319-99301-0}
\BIBentrySTDinterwordspacing

\bibitem{Kivela:2014:00}
M.~Kivel{\"a}, A.~Arenas, M.~Barthelemy, J.~P. Gleeson, Y.~Moreno, and M.~A.
  Porter, ``Multilayer networks,'' \emph{Journal of complex networks}, vol.~2,
  no.~3, pp. 203--271, 2014.

\bibitem{OMF:2022:00}
\BIBentryALTinterwordspacing
(2022, July). [Online]. Available:
  \url{https://openmodelingfoundation.github.io/about/}
\BIBentrySTDinterwordspacing

\bibitem{SE-Handbook-Working-Group:2015:00}
{SE Handbook Working Group}, \emph{Systems Engineering Handbook: A Guide for
  System Life Cycle Processes and Activities}.\hskip 1em plus 0.5em minus
  0.4em\relax International Council on Systems Engineering (INCOSE), 2015.

\bibitem{Hoyle:1998:00}
\BIBentryALTinterwordspacing
D.~Hoyle, \emph{{ISO 9000 pocket guide}}.\hskip 1em plus 0.5em minus
  0.4em\relax Oxford ; Boston: Butterworth-Heinemann, 1998. [Online].
  Available: \url{http://www.loc.gov/catdir/toc/els033/99163006.html}
\BIBentrySTDinterwordspacing

\bibitem{Farid:2006:IEM-C02}
\BIBentryALTinterwordspacing
A.~M. Farid and D.~C. McFarlane, ``{A Development of Degrees of Freedom for
  Manufacturing Systems},'' in \emph{IMS'2006: 5th International Symposium on
  Intelligent Manufacturing Systems: Agents and Virtual Worlds}, Sakarya,
  Turkey, 2006, pp. 1--6. [Online]. Available:
  \url{http://engineering.dartmouth.edu/liines/resources/Conferences/IEM-C02.pdf}
\BIBentrySTDinterwordspacing

\bibitem{Farid:2007:IEM-TP00}
\BIBentryALTinterwordspacing
A.~M. Farid, ``{Reconfigurability Measurement in Automated Manufacturing
  Systems},'' Ph.D. Dissertation, University of Cambridge Engineering
  Department Institute for Manufacturing, 2007. [Online]. Available:
  \url{http://engineering.dartmouth.edu/liines/resources/Theses/IEM-TP00.pdf}
\BIBentrySTDinterwordspacing

\bibitem{Farid:2008:IEM-J05}
\BIBentryALTinterwordspacing
A.~M. Farid and D.~C. McFarlane, ``{Production degrees of freedom as
  manufacturing system reconfiguration potential measures},'' \emph{Proceedings
  of the Institution of Mechanical Engineers, Part B (Journal of Engineering
  Manufacture) -- invited paper}, vol. 222, no. B10, pp. 1301--1314, 2008.
  [Online]. Available: \url{http://dx.doi.org/10.1243/09544054JEM1056}
\BIBentrySTDinterwordspacing

\bibitem{Farid:2008:IEM-J04}
\BIBentryALTinterwordspacing
A.~M. Farid, ``{Product Degrees of Freedom as Manufacturing System
  Reconfiguration Potential Measures},'' \emph{International Transactions on
  Systems Science and Applications -- invited paper}, vol.~4, no.~3, pp.
  227--242, 2008. [Online]. Available:
  \url{http://engineering.dartmouth.edu/liines/resources/Journals/IEM-J04.pdf}
\BIBentrySTDinterwordspacing

\bibitem{Farid:2015:ISC-J19}
\BIBentryALTinterwordspacing
------, ``{Static Resilience of Large Flexible Engineering Systems: Axiomatic
  Design Model and Measures},'' \emph{IEEE Systems Journal}, vol.~PP, no.~99,
  pp. 1--12, 2015. [Online]. Available:
  \url{http://dx.doi.org/10.1109/JSYST.2015.2428284}
\BIBentrySTDinterwordspacing

\bibitem{Farid:2016:ISC-BC06}
\BIBentryALTinterwordspacing
------, ``An engineering systems introduction to axiomatic design,'' in
  \emph{Axiomatic Design in Large Systems: Complex Products, Buildings \&
  Manufacturing Systems}, A.~M. Farid and N.~P. Suh, Eds.\hskip 1em plus 0.5em
  minus 0.4em\relax Berlin, Heidelberg: Springer, 2016, ch.~1, pp. 1--47.
  [Online]. Available: \url{http://dx.doi.org/10.1007/978-3-319-32388-6}
\BIBentrySTDinterwordspacing

\bibitem{Thompson:2020:02}
D.~Thompson, P.~Hegde, W.~C.~H. Schoonenberg, I.~Khayal, and A.~M. Farid, ``The
  hetero-functional graph theory toolbox,'' 2020.

\bibitem{farid:2021:00}
A.~M. Farid, D.~Thompson, P.~Hegde, and W.~Schoonenberg, ``A tensor-based
  formulation of hetero-functional graph theory,'' \emph{arXiv preprint
  arXiv:2101.07220}, 2021.

\bibitem{Thompson:2022:ISC-C80}
D.~Thompson and A.~M. Farid, ``Reconciling formal, multi-layer, and
  hetero-functional graphs with the hetero-functional incidence tensor,'' in
  \emph{IEEE Systems of Systems Engineering Conference}, Rochester, NY, 2022,
  pp. 1--6.

\bibitem{Amaral:2000:00}
L.~A.~N. Amaral, A.~Scala, M.~Barthelemy, and H.~E. Stanley, ``Classes of
  small-world networks,'' \emph{Proceedings of the national academy of
  sciences}, vol.~97, no.~21, pp. 11\,149--11\,152, 2000.

\bibitem{Albert:2000:00}
R.~Albert, H.~Jeong, and A.-L. Barab{\'a}si, ``Error and attack tolerance of
  complex networks,'' \emph{Nature}, vol. 406, no. 6794, pp. 378--382, Jul
  2000.

\bibitem{Barthelemy:2011:00}
M.~Barth{\'e}lemy, ``Spatial networks,'' \emph{Physics Reports}, vol. 499,
  no.~1, pp. 1--101, 2011.

\bibitem{Barabasi:1999:00}
A.-L. Barab{\'a}si and R.~Albert, ``Emergence of scaling in random networks,''
  \emph{science}, vol. 286, no. 5439, pp. 509--512, 1999.

\end{thebibliography}

\newpage
\newpage
\section*{Appendix}
  
\begin{table*}[b!]
\vspace{-0.1in}
\caption{This table presents the normalization factors for figures 3-7 for the regions NY, CA, TX, and the USA.}\label{Ta:Fig3-7N}
\vspace{-0.1in}
\begin{center}
\begin{tabular}{l c c c c}
& New York & California & Texas & United States\\\hline
Normalization Factor Figure 3  & 3.15e-4 & 1.37e-4 & 1.14e-4 & 7.93e-6\\\hline  
Normalization Factor Figure 4  & 1.30e-4 & 5.97e-5 & 5.07e-6 & 2.11e-6\\\hline 
Normalization Factor Figure 5  & 2.27e-4 & 1.17e-4 & 6.88e-5 & 5.60e-6\\\hline
Normalization Factor Figure 6  & 2.71e-5 & 1.52e-5 & 8.59e-6 & 9.25e-7\\\hline
Normalization Factor Figure 7A  & 2.29e-5 & 9.96e-6 & 6.99e-7 & 3.24e-7\\\hline  
Normalization Factor Figure 7B  & 2.29e-5 & 9.96e-6 & 6.99e-7 & 3.24e-7\\\bottomrule
\end{tabular}
\end{center}
\vspace{-0.1in}
\end{table*}

\begin{table*}[hb!]
\vspace{-0.1in}
\caption{This table presents the normalization factors for figures 8 and 9 for the regions NY, CA, TX, and the USA.}\label{Ta:Fig8-9N}
\vspace{-0.1in}
\begin{center}
\begin{tabular}{l c c c c c c c c}
& NY In & NY Out & CA In & CA Out & TX In & TX Out & USA In & USA Out\\\hline
Generate Electric Power from Water Energy   & 5.62e-3 & 5.62e-3 & 3.60e-3 & 3.60e-3 & 4.76e-2 & 4.76e-2 & 7.70E-4 & 7.70E-4\\\hline  
Generate Electric Power from Processed Gas  & 6.17e-3 & 6.17e-3 & 2.83e-3 & 2.83e-3 & 5.05e-3 & 5.05e-3 & 5.20E-4& 5.20E-4\\\hline 
Generate Electric Power from Processed Oil  & 4.17e-2 & 4.17e-2 & 0.125 & 0.125 & 0.33 & 0.33 & 1.63E-3 & 1.63E-3\\\hline
Generate Electric Power from Uranium        & 0.25 & 0.25 & 1 & 1 & 0.5 & 0.5 & 1.96E-2 & 1.96E-2\\\hline
Generate Electric Power from Coal           & 0.5 & 0.5 & 0.2 & 0.2 & 5.26e-2 & 5.26e-2 & 3.31E-3 & 3.31E-3\\\hline  
Generate Electric Power from Other          & 5.26e-2 & 5.26e-2 & 1.47e-2 & 1.47e-2 & 5.88e-2 & 5.88e-2 & 2.97E-3 & 2.97E-3\\\hline
Generate Electric Power from Wind Energy    & 2.56e-2 & 2.56e-2 & 9.43e-3 & 9.43e-3 & 7.69e-3 & 7.69e-3 & 2.10E-3 & 2.10E-3\\\hline
Generate Electric Power from Solar          & 5.26e-2 & 5.26e-2 & 2.39e-3 & 2.39e-3 & 5.56e-2 & 5.56e-2 & 1.56E-3 & 1.56E-3\\\hline
Consume Electric Power          & 4.23e-4 & 4.23e-4 & 1.79e-4 & 1.79e-4 & 1.65e-4 & 1.65e-4 & 1.56E-4 & 1.56E-4\\\hline
Compress processed gas          & 2.38e-2 & 2.38e-2 & 4.22e-3 & 4.22e-3 & 3.81e-3 & 3.81e-3 & 1.19E-2 & 1.19E-2\\\hline
Compress Syngas                 & 2.38e-2 & 2.38e-2 & 4.22e-3 & 4.22e-3 & 3.81e-3 & 3.81e-3 & 1.19E-2 & 1.19E-2\\\hline
Compress Raw Gas                & 2.38e-2 & 2.38e-2 & 4.22e-3 & 4.22e-3 & 3.81e-3 & 3.81e-3 & 1.19E-2 & 1.19E-2\\\hline
Import Processed Gas            & 3.94e-3 & 3.94e-3 & 1.16e-2 & 1.16e-2 & 7.11e-4 & 7.11e-4 & 2.45E-3 & 2.45E-3\\\hline
Import Syngas                   & 3.94e-3 & 3.94e-3 & 1.16e-2 & 1.16e-2 & 7.11e-4 & 7.11e-4 & 2.45E-3 & 2.45E-3\\\hline
Import Raw Gas                  & 3.94e-3 & 3.94e-3 & 1.16e-2 & 1.16e-2 & 7.11e-4 & 7.11e-4 & 2.45E-3 & 2.45E-3\\\hline
Export Processed Gas            & 3.94e-3 & 3.94e-3 & 1.16e-2 & 1.16e-2 & 7.11e-4 & 7.11e-4 & 2.45E-3 & 2.45E-3\\\hline
Import Crude Oil                & 1.35e-2 & 1.35e-2 & 1.02e-2 & 1.02e-2 & 4.63e-3 & 4.63e-3 & 2.13E-2 & 2.13E-2\\\hline
Export Crude Oil                & 1.35e-2 & 1.35e-2 & 1.02e-2 & 1.02e-2 & 4.63e-3 & 4.63e-3 & 2.13E-2 & 2.13E-2\\\hline
Import Processed Oil            & 1.35e-2 & 1.35e-2 & 1.02e-2 & 1.02e-2 & 4.63e-3 & 4.63e-3 & 2.13E-2 & 2.13E-2\\\hline
Export Processed Oil            & 1.35e-2 & 1.35e-2 & 1.02e-2 & 1.02e-2 & 4.63e-3 & 4.63e-3 & 2.17E-2 & 2.17E-2\\\hline
Import Liquid Biomass feedstock & 1.35e-2 & 1.35e-2 & 1.10e-2 & 1.10e-2 & 4.83e-3 & 4.83e-3 & 2.22E-2 & 2.22E-2\\\hline
Export Liquid Biomass feedstock & 1.35e-2 & 1.35e-2 & 1.10e-2 & 1.10e-2 & 4.83e-3 & 4.83e-3 & 2.22E-2 & 2.22E-2\\\hline
Import Coal                     & 0.25 & 0.25 & 0.125 & 0.125 & 1.82e-2 & 1.82e-2 & 2.66E-3 & 2.66E-3\\\hline
Export Coal                     & 0.33 & 0.33 & 0.17 & 0.17 & 6.67e-2 & 6.67e-2 & 0.05 & 0.05\\\hline
Process Raw Gas                 & 0 & 0 & 5.26e-2 & 5.26e-2 & 3.10e-3 & 3.10e-3 & 0.04 & 0.04\\\hline
Process Crude Oil               & 0 & 0 & 5.26e-2 & 5.26e-2 & 3.85e-2 & 3.85e-2 & 0.2 & 0.2\\\hline
Transport Electric Power        & 9.95e-5 & 9.95e-5 & 3.44e-5 & 3.44e-5 & 1.31e-4 & 1.31e-4 & 7.73E-5 & 7.73E-5\\\hline
Transport Processed Gas         & 1.52e-4 & 1.52e-4 & 6.43e-5 & 6.43e-5 & 1.63e-5 & 1.63e-5 & 2.07E-4 & 2.07E-4\\\hline
Transport Syngas                & 1.65e-4 & 1.65e-4 & 6.77e-5 & 6.77e-5 & 1.63e-5 & 1.63e-5 & 2.09E-4 & 2.09E-4\\\hline
Transport Raw Gas               & 1.65e-4 & 1.65e-4 & 6.77e-5 & 6.77e-5 & 1.63e-5 & 1.63e-5 & 2.07E-4 & 2.07E-4\\\hline
Transport Crude Oil             & 8.47e-3 & 8.47e-3 & 4.82e-4 & 4.82e-4 & 2.73e-4 & 2.73e-4 & 2.11E-3 & 2.11E-3\\\hline
Transport Processed Oil         & 3.23e-3 & 3.23e-3 & 5.92e-4 & 5.92e-4 & 5.37e-3 & 5.37e-3 & 1.05E-2 & 1.05E-2\\\hline
Transport Liquid Biomass Feedstock  & 0 & 0 & 0 & 0 & 0 & 0 & 0 & 0\\\hline
Transport Coal                  & 9.83e-5 & 9.83e-5 & 7.17e-5 & 7.17e-5 & 1.14e-3 & 1.14e-3 & 5.16E-4 & 5.16E-4\\\hline
Transport Water Energy          & 0 & 0 & 0 & 0 & 0 & 0 & 0 & 0\\\hline
Transport Other                 & 0 & 0 & 0 & 0 & 0.01 & 0.01 & 4.76E-2 & 4.76E-2\\\hline
Transport Solid Biomass Feedstock  & 0 & 0 & 0 & 0 & 0 & 0 & 0 & 0\\\hline
Total                           & 2.28e-5 & 2.28e-5 & 9.97e-6 & 9.97e-6 & 6.99e-7 & 6.99e-7 & 3.24E-7 & 3.24E-7\\\bottomrule
\end{tabular}
\end{center}
\vspace{-0.2in}
\end{table*}

\end{document}